\documentclass[letter paper,10pt]{article} 

\newcommand\authormark[1]{\textsuperscript{#1}}

\usepackage{amsmath}
\usepackage[a4paper, total={6in, 9.3in}]
{geometry}
\usepackage{graphicx}
\usepackage[colorlinks=true, bookmarks=false,citecolor=blue,urlcolor=blue]{hyperref}
\usepackage{natbib}
\usepackage{cleveref}

\usepackage{fancyhdr}% http://ctan.org/pkg/fancyhdr
\fancypagestyle{default}{
  \fancyhf{}% No header/footer
  \fancyfoot[R]{\thepage}% Right footer
  \renewcommand{\headrulewidth}{0pt}% No header rule
}

\begin{document}
\date{}
\title{Teaching CORnet Human fMRI Representations for Enhanced Model-Brain Alignment}
\maketitle
\vspace{-1.5cm}
\begin{center}
    Zitong Lu \authormark{1,*}, Yile Wang \authormark{2}
\end{center}

% Add Affiliation of Authors below
\begin{center}
$^{1}${Department of Psychology, The Ohio State University}\\
$^{2}${Department of Neuroscience, The University of Texas at Dallas}
\end{center}

\begin{center}
    *email: lu.2637@osu.edu
\end{center}
\pagestyle{default}
\renewcommand\headrulewidth{0pt}
\vspace{0.5cm}

\vspace{40pt}

\section*{Abstract}

Deep convolutional neural networks (DCNNs) have demonstrated excellent performance in object recognition and have been found to share some similarities with brain visual processing. However, the substantial gap between DCNNs and human visual perception still exists. Functional magnetic resonance imaging (fMRI) as a widely used technique in cognitive neuroscience can record neural activation in the human visual cortex during the process of visual perception. Can we teach DCNNs human fMRI signals to achieve a more brain-like model? To answer this question, this study proposed ReAlnet-fMRI, a model based on the SOTA vision model CORnet but optimized using human fMRI data through a multi-layer encoding-based alignment framework. This framework has been shown to effectively enable the model to learn human brain representations. The fMRI-optimized ReAlnet-fMRI exhibited higher similarity to the human brain than both CORnet and the control model in within- and across-subject as well as within- and across-modality model-brain (fMRI and EEG) alignment evaluations. Additionally, we conducted an in-depth analyses to investigate how the internal representations of ReAlnet-fMRI differ from CORnet in encoding various object dimensions. These findings provide the possibility of enhancing the brain-likeness of visual models by integrating human neural data, helping to bridge the gap between computer vision and visual neuroscience.

\vspace{20pt}

\noindent\textbf{Key words}: Brain-inspired Neural Networks; Object Recognition; fMRI; Neural Alignment; Human Brain-Like Model

\newpage

\section{Introduction}

Deep convolutional neural networks (DCNNs) in computer vision have rapidly advanced, achieving and even surpassing human performance in object recognition (\cite{Lecun2015}). These advancements not only drive the development of artificial intelligence (AI) but have also garnered significant interest from researchers in cognitive neuroscience. Numerous studies have observed that DCNNs not only internally represent basic visual features, such as orientation, position, shape, and texture (\cite{Yosinski2015, Zeiler2014}), but also capture important object attributes found in primate and human neuroimaging studies, such as animacy, spikiness, and real-world size (\cite{Cichy2014, Coggan2023, Khaligh-Razavi2018, Lu2023d, Yargholi2023, Konkle2013, Bao2020}). Additionally, some studies that directly compare CNNs with human brain representations have found that DCNNs exhibit hierarchical visual processing similar to the human visual system, both temporally and spatially (\cite{Cichy2016, Guclu2015, Kietzmann2019, Yamins2014, Lu2023c}). However, the presence of significant representational similarities does not imply that the two systems are highly alike. Both representational similarity analysis and regression-based measurements indicate that there remain substantial differences between DCNNs and the human brain in visual perception. 

Simply expanding the training dataset or increasing the number of layers within DCNNs cannot enhance model-brain alignment (\cite{Schrimpf2020}). Recent studies have attempted to optimize the architecture of DCNNs and have proposed a more brain-like recurrent DCNN called CORnet (\cite{Kubilius2018, Kubilius2019}). Although CORnet includes only four convolutional layers, it achieves the highest similarity to the brain as measured by Brain-Score (\cite{Schrimpf2020}), a platform for evaluating the similarity between models and primate brains. However, despite being considered the most brain-like visual system model to date, CORnet is still solely trained on images and has not been truly optimized using neural data. This raises the question: can we further refine CORnet, which was originally trained only on images, by using neural data to make it more brain-like?

To address this question, researchers have developed two main approaches to optimize DCNNs using neural data by refining the training process to bridge them closer to brain representations. One approach is the similarity-based method, which optimizes a similarity loss to make the model's representations more similar to neural activity (\cite{Dapello2023, Federer2020, Li2019, Pirlot2022}). This method has been used in studies to optimize DCNNs using neural activity from the mouse V1, or monkey V1 or IT. The other approach is the encoding-based method, which involves training DCNNs with an additional task to generate real brain neural activity (\cite{Safarani2021, Shao2024}), thus enabling the model to learn more brain-like internal representations. This method not only utilizes invasive neural activity from monkeys but also has recently been applied using non-invasive fMRI signals from humans to optimize ResNet. However, these studies have some limitations. Firstly, they often align a specific model layer with a specific brain region's neural activity, yet the correspondence between model layers and different brain regions is neither one-to-one nor well-understood. Secondly, these studies tend to focus more on whether the optimized model exhibits higher robustness rather than on model-brain alignment, which is the focus of our research.

A recent study proposed a novel multi-layer encoding alignment framework to effectively optimize CORnet using human EEG signals (\cite{Lu2024}). This human EEG-optimized model, called ReAlnet, was found to be significantly more similar to the human brain in terms of not only EEG but also fMRI and behavior. This may be one of the most effective methods for using neural data to optimize models, as it does not require specifying direct correspondences between different model layers and brain processing stages. Can we enhancing model-brain alignment by extending this framework to let CORnet learn human visual processing from fMRI data without specifying correspondences between its internal convolutional layers and different visual regions of the human brain?

In this study, we adapted CORnet to learn human fMRI representations using a modified multi-layer encoding alignment framework and proposed this fMRI-optimized model as ReAlnet-fMRI. We trained three personalized ReAlnet-fMRI models based on fMRI signals from three individual human subjects when they viewed natural images. We evaluated the model-brain alignment using within- and cross-subject within-modality fMRI data, as well as cross-subject across-modality EEG data. All evaluations suggest that ReAlnet-fMRIs exhibit enhanced model-brain alignment. Additionally, further internal representational analysis of both the purely image-trained CORnet and our fMRI-aligned ReAlnet-fMRIs revealed representational differences between these networks.

\section{Methods}

Here we first introduce the datasets we used in our study, including two fMRI datasets and one EEG dataset. Then, we describe the basic CORnet architecture and the training process of how we combined human fMRI signals to optimize CORnet to achieve our brain-aligned ReAlnet-fMRI models. Finally, we introduce the evaluation methods for measuring representational similarity between models and human brains and behaviors, and also the details of internal representational analysis on models we conducted to explore what ReAlnet-fMRI learned from human fMRI representations.

\subsection{Human Neuroimaging Datasets}

\textbf{fMRI dataset for model training}: The fMRI data originates from \cite{Shen2019}. This \textit{Shen fMRI dataset} recorded human brain fMRI signals from three subjects while they focused on the center of the screen viewing natural images from ImageNet. We applied the training set from \textit{Shen fMRI dataset}, which comprises fMRI signals of each subject viewing 1,200 images (from 150 object categories, 8 images per category) with each image being viewed 5 times and averaged the fMRI signals across the repeated trials to obtain more stable brain activity for each image observation to train our ReAlnet-fMRIs (1,200 samples for training).

Here, we selected the voxels from the entire visual cortex to obtain fMRI signals, and we applied principal component analysis (PCA) based on individual training data to reduce the total number of voxels in the visual cortex to 1,024 feature dimensions. Consequently, the training data corresponding to each human subject consisted of 1,200 samples $\times$ 1,024 features.

\noindent\textbf{fMRI dataset for model test (within-modality \& within-subject):} To evaluate whether ReAlnet-fMRIs shows higher similarity to human fMRI representations, we applied the test set from \textit{Shen fMRI dataset} to test the within-modality and within-subject model-fMRI similarity. This test set comprises fMRI signals of same three subjects viewing 50 images (from 50 object categories from ImageNet but different from 150 training categories), 40 artificial shape images, and 10 alphabetical letter images with each image being viewed 24, 20, and 12 times respectively.

Similar to what we did for the training set, We averaged the fMRI signals across the repeated trials to obtain more stable brain activity for each image observation. For the testing, we extracted signals from five regions-of-interest (ROIs) for subsequent comparison of model-fMRI similarity: V1, V2, V3, V4, and the lateral occipital complex (LOC).

\noindent\textbf{fMRI dataset for model test (within-modality \& across-subject):} Although our ReAlnet-fMRIs were trained on individual fMRI signals, we also would like to whether these models learn more general brain representations across individuals than the original purely image-trained CORnet. Thus we selected another fMRI dataset (\cite{Horikawa2017}), \textit{Horikawa fMRI dataset}, which included fMRI signals from five different subjects viewing natural images. Here, we used the test set of \textit{Horikawa fMRI dataset} to evaluate the within-modality but across-subject model-fMRI similarity. This test set comprises fMRI signals of five subjects viewing 50 images (as same as images used in \textit{Shen fMRI dataset}'s test set) with each image being repeated 35 times. We also averaged the repeated trials and extracted signals from V1, V2, V3, V4, and LOC to calculate the model-fMRI similarity.

\noindent\textbf{EEG dataset for model test (across-modality \& across-subject):} To further confirm that our ReAlnet-fMRIs learn more general human brain representations instead of just human fMRI representations, we need to apply across-modality human neuroimaging data to conduct the model-brain alignment evaluation. Here we obtained human EEG data from an EEG open dataset, \textit{THINGS EEG2 dataset} (\cite{Gifford2022}), including EEG data from 10 healthy human subjects in a rapid serial visual presentation (RSVP) paradigm. Stimuli were images sized 500 $\times$ 500 pixels from \textit{THINGS dataset} (\cite{Hebart2019}), which consists of images of objects on a natural background from 1854 different object concepts. We applied the test set in \textit{THINGS EEG2 dataset} to evaluate the across-modality and also across-subject model-EEG similarity. In this test set, each subject completed 16,000 trials with 200 images from 200 object concepts and 80 repeated trials per images. Subjects viewed one image per trial (100ms).

EEG data were collected using a 64-channel EASYCAP and a BrainVision actiCHamp amplifier. We used already pre-processed data from 17 channels (O1, Oz, O2, PO7, PO3, POz, PO4, PO8, P7, P5, P3, P1, Pz, P2) overlying occipital and parietal cortex. We re-epoched EEG data ranging from stimulus onset to 200ms after onset with a sample frequency of 100Hz. Thus, the shape of our EEG data matrix for each trial is 17 channels $\times$ 20 time points. Similar to fMRI, we averaged all the repeated trials for each image to obtain more stable EEG signals.

\subsection{Model Architecture and Training}

\noindent\textbf{Basic architecture of ReAlnet-fMRI:} We have chosen the state-of-the-art CORnet-S model (\cite{Kubilius2018, Kubilius2019}) as the foundational architecture for ReAlnet-fMRI. Both CORnet and ReAlnet consist of four visual layers (V1, V2, V4, and IT) and a category decoder layer. Layer V1 performs a 7 $\times$ 7 convolution with a stride of 2, followed by a 3 $\times$ 3 max pooling with a stride of 2, and another 3 $\times$ 3 convolution. Layer V2, V4, and IT each perform two 1 $\times$ 1 convolutions, a bottleneck-style 3 $\times$ 3 convolution with a stride of 2, and a 1 $\times$ 1 convolution. Apart from the initial Layer V1, the other three visual layers include recurrent connections, allowing outputs of a certain layer to be passed through the same layer several times (twice in Layer V2 and IT, and four times in Layer V4). 

\noindent\textbf{Image-to-fMRI encoding-based alignment framework:} In addition to the original CORnet structure, we have added an fMRI generation module designed to construct an image-to-fMRI encoding model for generating human fMRI signals from human visual cortex (\Cref{Figure1}A). Each visual layer is connected to a nonlinear \textit{N} $\times$ 128 layer-encoder (Enc-V1, Enc-V2, Enc-V4, and Enc-IT correspond to Layer V1, V2, V4, and IT) that processes through a fully connected network with a ReLU activation. These four layer-encoders are then directly concatenated to form an \textit{N} $\times$ 512 Multi-Layer Visual Encoder, which is subsequently connected to an \textit{N} $\times$ 1024 fMRI encoder through a linear layer to generate the predicted fMRI signals. Here \textit{N} is the batch size. Therefore, we aim for ReAlnet-fMRI to not only perform the object classification task but also to generate human fMRI signals through the fMRI generation module to learn human fMRI representations when the human subject views the certain image. During this process of predicting huamn fMRI activity, ReAlnet-fMRI’s visual layers are poised to effectively extract features more aligned with neural representations.

Similar to the study of ReAlnet (\cite{Lu2024}), the training loss $\mathcal{L}^A$ of this modified alignment framework consists of a classification loss and a generation loss with a parameter $\beta$ that determines the relative weighting:

\begin{equation}
\mathcal{L}^A=\mathcal{L}^C+ \beta \cdot \mathcal{L}^G
\end{equation}

$\mathcal{L}^C$ represents the standard categorical cross entropy loss for model predictions on ImageNet labels:

\begin{equation}
    \mathcal{L}^C=-\sum_{i=1}^N y_i log(p_i)
\end{equation}

Here, $y_i$ represents the $i$-th image, and $p_i$ represents the probability that model predicts the $i$-th image belongs to class $i$ out of 1000 categories. However, some images in \textit{Shen fMRI dataset} were not included in 1,000 categories in ImageNet 1,000 category version. Therefore, we adopt the same strategy as in previous studies (\cite{Dapello2023, Lu2024}), using the labels obtained from the ImageNet pre-trained CORnet without neural alignment as the true labels to stabilize the classification performance of ReAlnet-fMRI.

$\mathcal{L}^G$ represents the generation loss including a mean squared error (MSE) loss and a contrastive loss between the generated and real fMRI signals. This contrastive loss is calculated based on the dissimilarity (1 minus Spearman correlation coefficient) between generated and real signals, aiming to bring the generated signals from the same image (positive pairs) closer to the corresponding real human fMRI signals and make the generated signals from different images (negative pairs) more distinct. $\mathcal{L}^G$ is calculated as followed:

\begin{equation}
    \mathcal{L}^G = \frac{1}{N}\sum_{i=1}^N(S_i - \hat{S_i})^2 + \frac{1}{N}\sum_{i=1}^N[1-\rho(S_i, \hat{S_i})] \\ - \frac{1}{N(N-1)}\sum_{i=1}^N\sum_{j=1, j\ne i}^N[1- \rho(S_i, \hat{S_j})]
\end{equation}

Here, $S_i$ and $\hat{S}_i$ represent the generated and real fMRI signals corresponding to the $i$-th image.

\noindent\textbf{Training procedures:} We trained 3 individual ReAlnet-fMRI independently based on 3 human subjects' fMRI data in \textit{Shen fMRI dataset} independently. Each network was trained to minimize the alignment loss including both classification and generation losses with a static loss weight $\beta$ of 10, 20, 30, 40 or 50 and a static training rate of 0.00002 for 5 epochs using the Adam optimizer. We used a batch size of 16, meaning the contrastive loss computed dissimilarities of 256 pairs for each gradient step. 

Additionally, the absence of correct ImageNet labels for many images in \textit{Shen fMRI dataset} should indeed decrease the category classification performance on ImageNet. To better control, we trained a control model with $\beta$ = 0, called Control. We tested the classification accuracy on ImageNet and behavior similarity on Brain-Score (See \textit{Model-Brain Similarity Measurement} section) of CORnet, Control, and ReAlnet-fMRIs at different $\beta$ values (\Cref{FigureS1}). Although the classification accuracy of ReAlnet-fMRIs decreased compared with CORnet and there was a slight tendency to decrease as $\beta$ increases, but hardly changed compared with Control. Also, there was no significant different on behavioral similarity between CORnet and ReAlnet-fMRIs when beta is larger than 10, and all ReAlnet-fMRIs showed higher behavioral similarity than Control. In the main text, we show the results of ReAlnet-fMRIs with $\beta$ = 40 (See the results of ReAlnet-fMRIs with $\beta$ = 10, 20, 30, 50 in Supplementary \Cref{FigureS2,FigureS3,FigureS4,FigureS5,FigureS6,FigureS7,FigureS8,FigureS9,FigureS10,FigureS11,FigureS12,FigureS13}).

\subsection{Model-Brain Similarity Measurement}

Representational similarity analysis (RSA) (\cite{Kriegeskorte2008a}) is used for representational comparisons between models and human brain activity. First, we computed representational dissimilarity matrices (RDMs) for models and human fMRI or EEG signals. Then, we calculated Spearman correlation coefficients between model RDMs and human neural RDMs. All RSA analyses were implemented based on NeuroRA toolbox (\cite{Lu2020}).

\noindent\textbf{Model-fMRI similarity:} To evaluate the within-subject model-fMRI similarity, the shape of each RDM (natural images, artificial shape images, or alphabetical letter images) is 50 $\times$ 50, 40 $\times$ 40, or 10 $\times$ 10 in \textit{Shen fMRI dataset} test set. For fMRI RDMs, we calculated 1 minus Pearson correlation coefficient between voxel-wise activation patterns corresponding to any two images as the dissimilarity index in the RDM for each ROI and each subject. For model RDMs, we input 50 natural images, 40 artificial shape images, and 10 alphabetical letter images respectively into each model and obtained latent features from each visual layer. Then, we constructed each layer’s RDM by calculating the dissimilarity using 1 minus Pearson correlation coefficient between flattened vectors of latent features corresponding to any two images. To compare the representations, we calculated the Spearman correlation coefficient as the similarity index between layer-by-layer model RDMs and neural fMRI RDMs corresponding to different ROIs, assigning the final similarity for a certain brain region as the highest similarity result across model layers due to the lack of a clear correspondence between different model layers and brain regions. Similarity, to evaluate the across-subject model-fMRI similarity, the only difference was that we obtained 50 $\times$ 50 fMRI RDMs corresponding to 5 ROIs for each subject in \textit{Horikawa fMRI dataset} test set. 

\noindent\textbf{Model-EEG similarity:} To evaluate the across-modality and across-subject model-EEG similarity, the shape of each RDM is 200 $\times$ 200, corresponding to 200 images in THINGS EEG2 test set. For EEG RDMs, we applied timepoint-by-timepoint classification-based EEG decoding and used decoding accuracy between two image conditions as the dissimilarity index to construct EEG RDM for each timepoint and each subject. For model RDMs, we input 200 images into each model and obtained the final layer-by-layer model RDMs. To compare the representational similarity temporally, we calculated the Spearman correlation coefficient between layer-by-layer model RDMs and timepoint-by-timepoint neural EEG RDMs.

\noindent\textbf{Model-behavior similarity:} We measured the model-behavior similarity based on Brain-Score, which is a framework evaluating how similar the model is to the primate visual system (\cite{Schrimpf2020}). Here, we applied the behavioral benchmarks (including "Rajalingham2018public-i2n" assessing the ability of recognizing core objects from visual images, even with various changes in position, size, viewing angle, and background of the objects (\cite{Rajalingham2018}) and "Geirhos2021-error\_consistency" measuring the similarity of errors made by ANN and human (\cite{Geirhos2021})) from Brain-Score platform. We obtained the behavior similarity of both CORnet and ReAlnet-fMRI with different $\beta$ values. For more detailed information about the behavioral benchmarks in Brain-Score, please refer to the original papers (\cite{Schrimpf2020, Rajalingham2018, Geirhos2021}).

\subsection{Model Internal Representational Analysis}

To assess the model's encoding of different object features and explore which object dimension the fMRI-optimized ReAlnet-fMRI encodes more strongly or weakly compared to CORnet, we applied 49 object space dimensions from THINGS (\cite{Hebart2020}), where each object concept can be embedded along these 49 dimensions. Here, our analysis is based on the 200 images in the test set of \textit{THINGS EEG2 dataset} (these images were not involved in the model's training). We employed an RDM-based partial Spearman correlation method for the analysis. Specifically, we first computed the RDM for the IT layer (which contained more higher-level information) of each model and 49 feature RDMs based on 200 images (calculating the absolute value differences in feature encoding strength on the same dimension between every two images as dissimilarity). Then, we computed the partial correlation between the model RDM and each feature RDM, regressing our the other 48 feature RDMs, and calculated the square of the partial correlation coefficient as the explained variance of the model by that object dimension.

\section{Results}

\subsection{Within-Modality \& Within-Subject Model-fMRI Similarity}

\begin{figure}[ht!]
\vskip 0in
\begin{center}
\centerline{\includegraphics[width=\columnwidth]{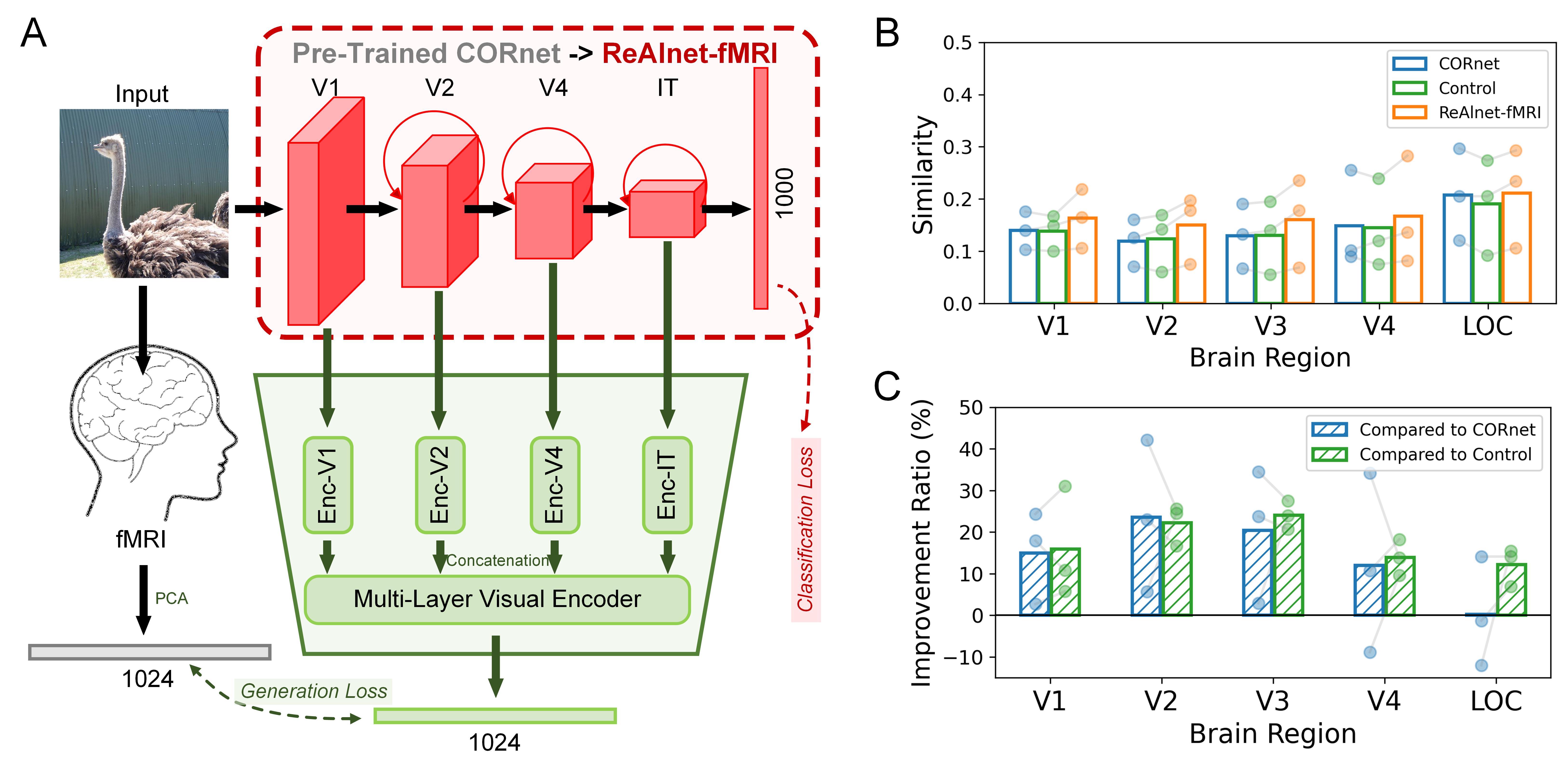}}
\caption{Human fMRI-optimized ReAlnet-fMRI as a more human brain-like vision model. (A) An overview of ReAlnet-fMRI alignment framework. Adding an additional multi-layer encoder to an ImageNet pre-trained CORnet-S, the outputs contain the category classification results and the generated fMRI signals with two losses, a classification loss and a generation loss. (B) Within-subject representational similarity between models (CORnet, Control, and ReAlnet-fMRIs) and human fMRI on natural images. (C) Similarity Improvement ratio of within-subject model-fMRI similarity on natural images of ReAlnet-fMRIs compared to other two models. Each circle dot indicates an individual ReAlnet-fMRI.}
\label{Figure1}
\end{center}
\vskip -0.2in
\end{figure}

First, we would like to confirm that training the model using fMRI signals of humans viewing natural images can make the model's representations more similar to human fMRI representation at the within-subject level. Based on the analysis of the 50 natural images in the test set from \textit{Shen fMRI dataset}, we calculated the similarity between (1) fMRI RDMs based on 5 different ROIs and RDMs of CORnet based on 4 layers, (2) their fMRI RDMs and RDMs of Control, (3) their fMRI RDMs and RDMs of the subject-matched ReAlnet-fMRI. The results showed that ReAlnet-fMRIs were more similar to human fMRI representations across various visual brain regions compared to both CORnet and Control (\Cref{Figure1}B).

We also analyzed  the improvement ratio of similarity (compared to CORnet: (ReAlnet-fMRI - CORnet)/CORnet; compared to Control: (ReAlnet-fMRI - Control)/Control) (\Cref{Figure1}C). The average improvement ratio exceeded 10\%, with the highest improvement ratio reaching 43\%. Although there were 1-2 instances where ReAlnet-fMRIs exhibited lower similarity to human V4 and LOC representations compared to CORnet, they were still higher than Control. These few instances of negative improvement are likely due to the lack of ImageNet category labels for some images during the training of ReAlnet-fMRI. Nonetheless, our image-to-fMRI encoding-based alignment framework demonstrated a significant capability to enhance model-fMRI alignment.

\begin{figure}[ht!]
\vskip 0in
\begin{center}
\centerline{\includegraphics[width=0.8\columnwidth]{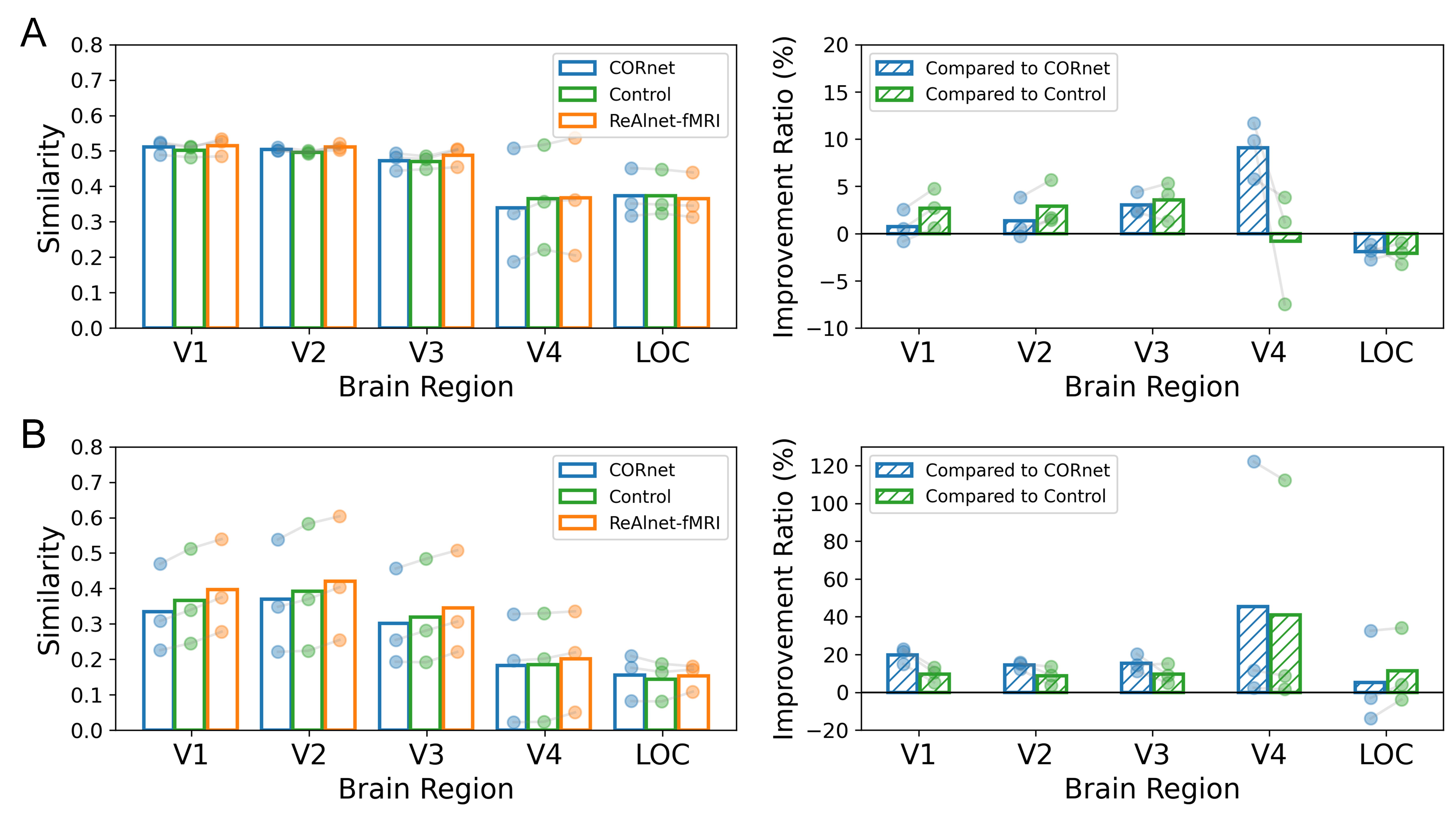}}
\caption{Within-subject model-fMRI similarity and similarity improvement ratio on (A) artificial shape images and (B) alphabetical letter images. Each circle dot indicates an individual ReAlnet-fMRI.}
\label{Figure2}
\end{center}
\vskip -0.2in
\end{figure}

Furthermore, can ReAlnet-fMRI, which is trained to learn fMRI representations corresponding to natural images, also learn brain representations when humans view other categories of images? To investigate this, we selected two other parts of data from the test set of \textit{Shen fMRI dataset}, which includes 40 artificial shape images and 10 alphabetical letter images. The results showed that for simple shape and letter images, ReAlnet-fMRIs still exhibited higher similarity to human brain representations (\Cref{Figure2}A-B). However, in V4 and LOC brain regions, we observed instances where ReAlnet-fMRIs had lower similarity compared to CORnet and/or Control. This might be primarily due to the fact that in generalizing across different image categories, ReAlnet-fMRIs reflect more brain processing of low-level visual information according to images only including shape or letter with more low-level features.

\subsection{Within-Modality \& Across-subject Model-fMRI Similarity}

\begin{figure}[ht!]
\vskip 0in
\begin{center}
\centerline{\includegraphics[width=\columnwidth]{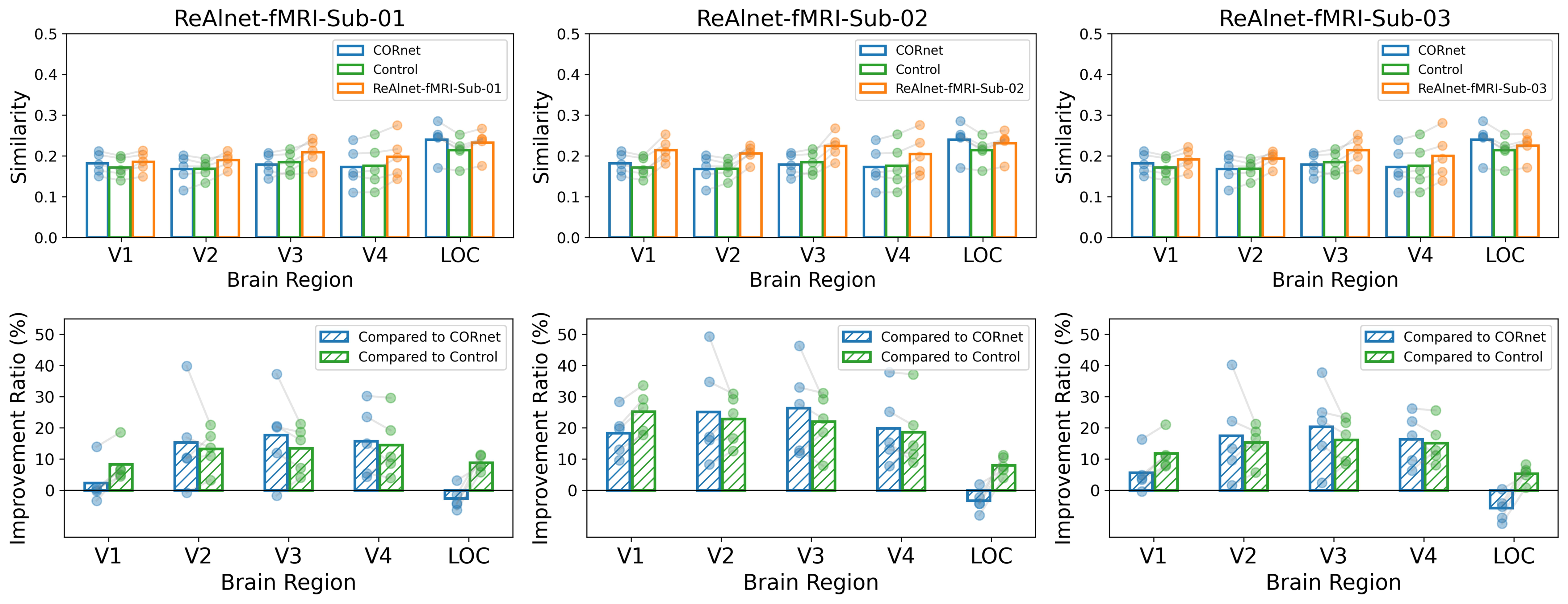}}
\caption{Across-subject model-fMRI similarity and similarity improvement ratio. Each circle dot indicates a subject from \textit{Horikawa fMRI dataset}.}
\label{Figure3}
\end{center}
\vskip -0.2in
\end{figure}

To further confirm that our individually fMRI-optimized ReAlnet-fMRIs learn not only individual-specific but also more general brain representations across individuals, compared to the original purely image-trained CORnet, we conducted the across-subject model-fMRI similarity analysis on the test set from \textit{Horikawa fMRI dataset}. The results were consistent with the within-subject (subject-matched) similarity results from \textit{Shen fMRI dataset} above, showing that three ReAlnet-fMRI models exhibited higher similarity to the fMRI representations of all five subjects in \textit{Horikawa fMRI dataset} (\Cref{Figure3}).

\subsection{Across-Modality \& Across-Subject Model-EEG Similarity}

\begin{figure}[ht!]
\vskip 0in
\begin{center}
\centerline{\includegraphics[width=\columnwidth]{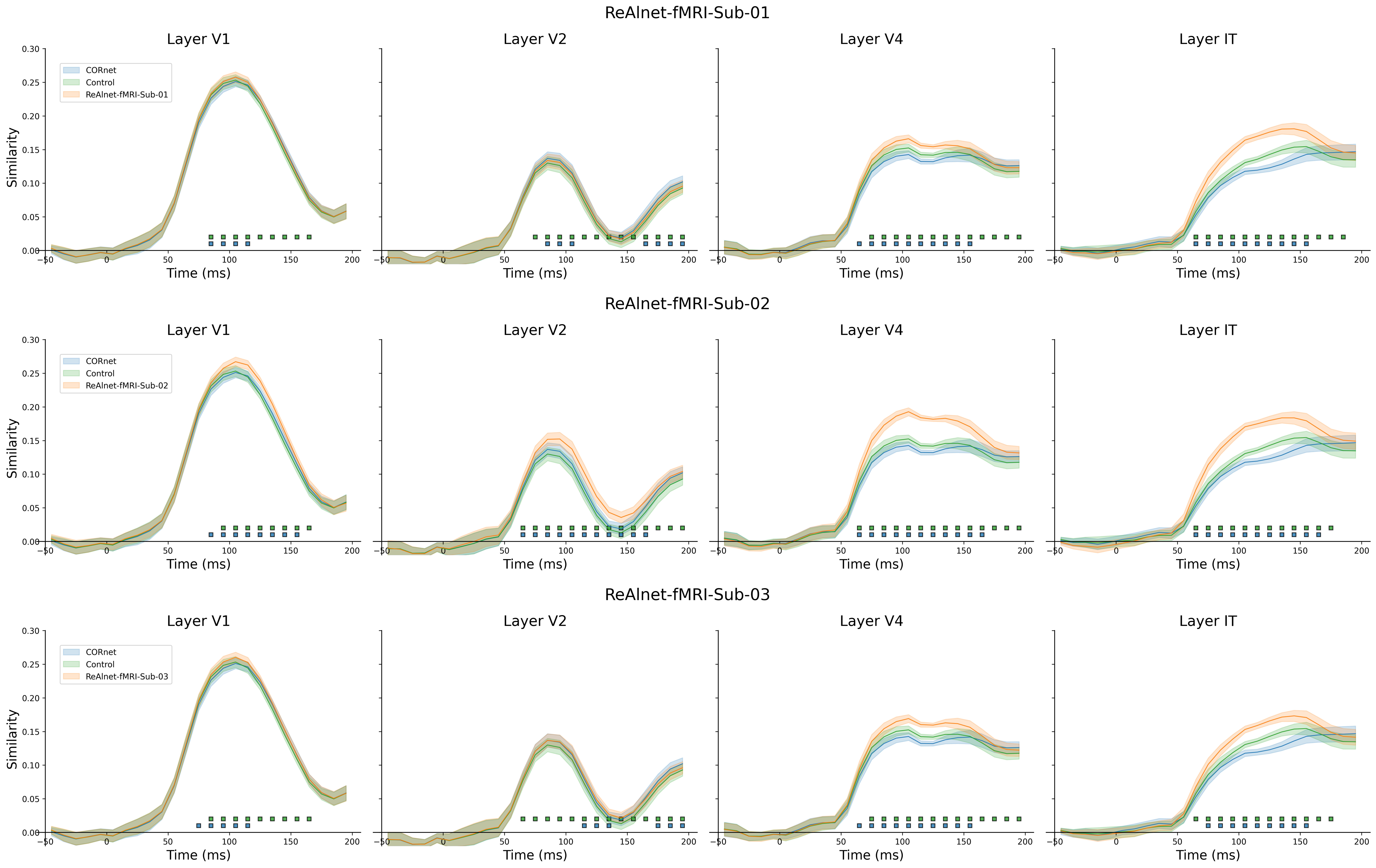}}
\caption{Across-subject temporal model-EEG similarity. Blue and green square dots with black outlines at the bottom indicate the timepoints where ReAlnet-fMRI vs. CORnet and ReAlnet-fMRI vs. Control were significantly different ($p<.05$). Shaded area reflects ±SEM.}
\label{Figure4}
\end{center}
\vskip -0.2in
\end{figure}

Although we observed that ReAlnet-fMRIs exhibit higher similarity to human fMRI representations, it is important to note that ReAlnet-fMRIs are trained based on fMRI signals. This raises the question: are ReAnet-fMRIs learning just the fMRI representations, or are they capturing broader human brain representations in visual perception? If it is the latter, we should be able to observe that ReAlnet-fMRIs also show higher similarity to across-modality human EEG representations compared to CORnet. To test this, we conducted an across-modality and across subject model-EEG similarity analysis using the EEG data from 10 subjects in the test set of \textit{THINGS EEG2 dataset}. The results showed that ReAlnet-fMRIs have significantly higher similarity to human EEG neural dynamics across all four visual layers than both CORnet and Control without human neural alignment (\Cref{Figure4}). This indicates that ReAlnet-fMRIs learn broader, cross-modality human brain representations from human fMRI signals, not just fMRI representations.

\subsection{Internal Representational Analysis}

\begin{figure}[ht!]
\vskip 0in
\begin{center}
\centerline{\includegraphics[width=\columnwidth]{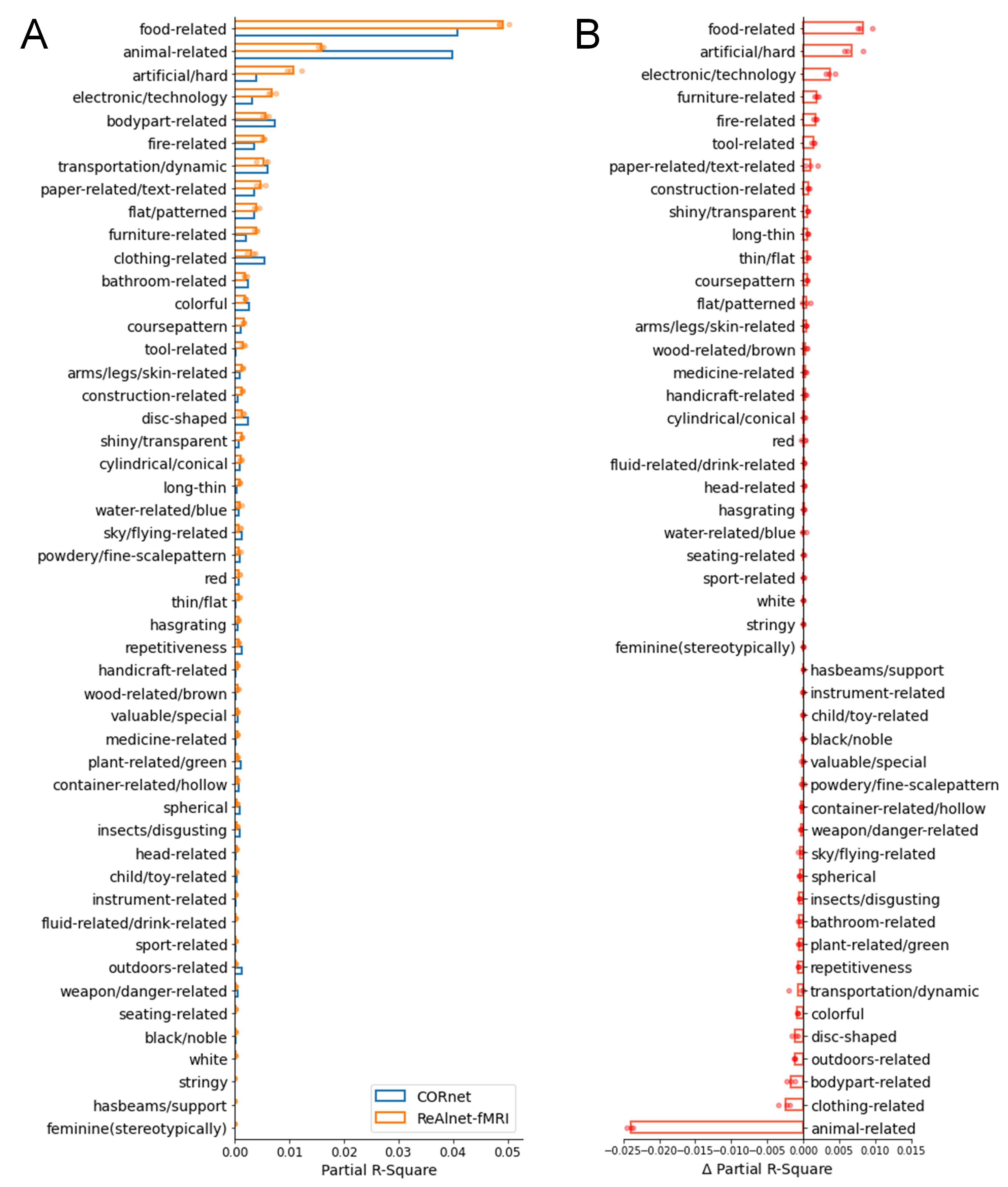}}
\caption{Internal representations in ReAlnet-fMRIs and CORnet. (A) Partial r-square of each object dimension in ReAlnet-fMRIs and CORnet. (B) The difference of partial r-square between ReAlnet-fMRIs and CORnet. Each circle dot indicates an individual ReAlnet-fMRI.}
\label{Figure5}
\end{center}
\vskip -0.2in
\end{figure}

Given that our image-to-fMRI encoding-based alignment framework can indeed help CORnet learn human visual representations from human fMRI signals and improve model-brain alignment, the next question is: what are the differences in visual information encoding between the fMRI-optimized ReAlnet-fMRIs and CORnet? Specially, what aspects of information encoding are enhanced by the brain fMRI data in ReAlnet-fMRIs? We conducted an in-depth analysis of the IT layer of both ReAlnet-fMRIs and CORnet to examine their encoding of 49 object dimensions. These internal representational analysis results showed that both ReAlnet-fMRIs and CORnet encode a wide range of object visual features, with stronger encoding observed in dimensions of food-related, animal-related, artificial/hard, electronic/technology, and body/parts (\Cref{Figure5}A). Further calculating the differences between ReAlnet-fMRIs and CORnet, we observed that ReAlnet-fMRIs exhibit stronger processing in food-related, artificial/hard, and electronics/technology dimensions, while showing weaker processing in animal-related dimensions compared to CORnet. It is important to note that these differences do not imply that ReAlnet-fMRIs or CORnet fail to encode certain object dimensions. Instead, they suggest that models trained solely on image data may not capture as much information related to food, artificial objects, and electronics as models optimized with human neural data. Conversely, they might overly capture animal-related information. More importantly, these results also indicate that the brain encompasses more food-related, artificial/hard, and electronic/technology information, which is effectively captured and learned by our ReAlnet-fMRIs.

\section{Discussion}

Our study aimed to teach the current state-of-the-art vision model, CORnet, human fMRI representations by training the model based on an image-to-fMRI encoding-based alignment framework. These human fMRI-optimized models, ReAlnet-fMRIs, enhanced the model's alignment with human brain representations. We evaluated this at different levels, and the results from multiple experiments based on multiple neuroimaging datasets confirmed that this alignment improvement was observed not only in within-modality and within-subject model-fMRI comparisons but also in across-subject model-fMRI and across-modality model-EEG comparisons. This indicates that ReAlnet-fMRIs, to some extent, captured the way how human brains process visual information and utilized brain fMRI data to optimize their weights in a brain-like manner within our alignment training process. This optimization process demonstrated strong generaliztion across images, image categories, brain imaging modalities, and individuals, ultimately resulting in a comprehensive enhance model-brain alignment.

Additionally, we analyzed the differences in internal representations between fMRI-optimized ReAlnet-fMRIs and the purely image-trained CORnet. Interestingly, we observed stronger encoding of information related to food and other several features in ReAlnet-fMRIs. These differences highlight the distinct ways the brain processes visual information compared to models trained purely on image-based patter recognition.

Why does CORnet exhibit weaker encoding of such information? On one hand, CORnet is trained on the 1000-category classification task of ImageNet, an object recognition benchmark that does not specifically require the model to capture higher-level information, such as food-related or artificial/hard object information. Perhaps, we can improve DCNNs by augmenting the original ImageNet 1000 classification task with additional tasks that require more fine-grained recognition, but are common in everyday human experience, such as food and tool categorizations. This may enhance the model's encoding of these important features. On the other hand, information such as food is often learned through multiple sensory systems in real life, including taste, smell, motion, and even language, rather than through vision alone. CORnet, being a vision-only model, cannot learn these comprehensive encoding patterns as the human brain does. Many human fMRI studies have discovered the evidence of food-related and artificial object information encoding in human brains (\cite{Cichy2014, Jain2023, Khosla2022}), making it reasonable that ReAlnet-fMRIs, optimized with human fMRI signals, exhibits stronger encoding of such information and more brain-like representations.

To achieve more brain-like visual models, besides approaches like our current work that directly use neural data for model optimization (\cite{Dapello2023, Federer2020, Li2019, Pirlot2022, Lu2024, Shao2024}), there are other strategies as well. These include modifying the model's architecture by constructing dual-way pathways models similar to the brain's visual pathways (\cite{Bai2017, Choi2023, Han2022, Han2023, Sun2017}), adding feedback pathways (\cite{Konkle2023}), incorporating topographic constraints (\cite{Finzi2022, Lee2020, Lu2023e, Margalit2023}), or changing the model's training tasks by using self-supervised training or training the model in a richer 3D environment (\cite{Konkle2022, Prince2023}). These different approaches to realizing brain-inspired AI are not mutually exclusive; in face, they may complement each other. Current researchers from both computer science and neuroscience are exploring various angles to word toward this goal.

Current research also has some limitations. From a data perspective, the primary limitations of our current study stem from (1) the relatively smaller sample size of neural datasets compared to image datasets with vast samples, and (2) the lack of shared labels between different datasets, such as the absence of some ImageNet category labels for images used in human neuroimaging studies. These limitations restrict further enhancement of ReAlnet-fMRI’s similarity to the human brain and reduce its classification performance on ImageNet. From a technical perspective, future research may need to focus on (1) more effectively learning the alignment of models with the human brain using small-sample neural data, and (2) employing self-supervised or unsupervised learning methods that do not require category labels for model training.

In future work, we hope to extend this multi-layer alignment framework to broader domains, such as language models, auditory models, and even cross-modal models. Certainly, these extended applications will also necessitate corresponding neural data collection efforts. Additionally, beyond focusing on model-brain alignment, it is worth testing whether models optimized using neural data show performance improvements in tasks where traditional models perform poorly. For instance, testing the robustness of models under different noise conditions to see if there is an improvement (some studies have already found brain-aligned models to have higher adversarial robustness) (\cite{Dapello2023, Shao2024}), and evaluating whether the models exhibit stronger abstraction and generalization capabilities.

In summary, we utilized a neural alignment framework capable of synchronously optimizing multiple layers of CORnet and trained three individualized ReAlnet-fMRI models based on this framework. Through a series of detailed and comprehensive evaluations, we demonstrated that ReAlnet-fMRIs are more human brain-like visual models that exhibit significantly enhance model-brain alignment. In addition, the training process of learning fMRI representations enabled ReAlnet-fMRis to show stronger encoding in object dimensions such as food-related features compared to CORnet. We hope that our work helps bridge the gap between AI in computer vision and human visual neuroscience, thereby achieving more brain-like intelligence and understanding the brain from an AI perspective.

\section*{Acknowledgment}

This research did not receive any specific grant from funding agencies in the public, commerical, or not-for-profit sectors. We thank Julie Golomb for helpful discussions. We thank the Ohio Supercomputer Center and Georgia Stuart for providing the essential computing resources and support.

% \bibliographystyle{plain}
% \printbibliography
\bibliographystyle{apalike}
\bibliography{references}

\begin{thebibliography}{}

\bibitem[Bai et~al., 2017]{Bai2017}
Bai, S., Li, Z., and Hou, J. (2017).
\newblock {Learning two-pathway convolutional neural networks for categorizing scene images}.
\newblock {\em Multimedia Tools and Applications}, 76(15):16145--16162.

\bibitem[Bao et~al., 2020]{Bao2020}
Bao, P., She, L., McGill, M., and Tsao, D.~Y. (2020).
\newblock {A map of object space in primate inferotemporal cortex}.
\newblock {\em Nature}, 583(7814):103--108.

\bibitem[Choi et~al., 2023]{Choi2023}
Choi, M., Han, K., Wang, X., Zhang, Y., and Liu, Z. (2023).
\newblock {A Dual-Stream Neural Network Explains the Functional Segregation of Dorsal and Ventral Visual Pathways in Human Brains}.
\newblock {\em Advances in Neural Information Processing Systems (NeurIPS)}, 36.

\bibitem[Cichy et~al., 2016]{Cichy2016}
Cichy, R.~M., Khosla, A., Pantazis, D., Torralba, A., and Oliva, A. (2016).
\newblock {Comparison of deep neural networks to spatio-temporal cortical dynamics of human visual object recognition reveals hierarchical correspondence}.
\newblock {\em Scientific Reports}, 6(1):1--13.

\bibitem[Cichy et~al., 2014]{Cichy2014}
Cichy, R.~M., Pantazis, D., and Oliva, A. (2014).
\newblock {Resolving human object recognition in space and time}.
\newblock {\em Nature Neuroscience}, 17(3):455--462.

\bibitem[Coggan and Tong, 2023]{Coggan2023}
Coggan, D.~D. and Tong, F. (2023).
\newblock {Spikiness and animacy as potential organizing principles of human ventral visual cortex}.
\newblock {\em Cerebral Cortex}, 33(13):8194--8217.

\bibitem[Dapello et~al., 2023]{Dapello2023}
Dapello, J., Kar, K., Schrimpf, M., Geary, R.~B., Ferguson, M., Cox, D.~D., and DiCarlo, J.~J. (2023).
\newblock {Aligning Model and Macaque Inferior Temporal Cortex Representations Improves Model-to-Human Behavioral Alignment and Adversarial Robustness}.
\newblock {\em International Conference on Learning Representations (ICLR)}, 12.

\bibitem[Federer et~al., 2020]{Federer2020}
Federer, C., Xu, H., Fyshe, A., and Zylberberg, J. (2020).
\newblock {Improved object recognition using neural networks trained to mimic the brain's statistical properties}.
\newblock {\em Neural Networks}, 131:103--114.

\bibitem[Finzi et~al., 2022]{Finzi2022}
Finzi, D., Margalit, E., Kay, K., Yamins, D. L.~K., and Grill-Spector, K. (2022).
\newblock {Topographic DCNNs trained on a single self-supervised task capture the functional organization of cortex into visual processing streams}.
\newblock {\em NeurIPS 2022 Workshop SVRHM}.

\bibitem[Geirhos et~al., 2021]{Geirhos2021}
Geirhos, R., Narayanappa, K., Mitzkus, B., Thieringer, T., Bethge, M., Wichmann, F.~A., and Brendel, W. (2021).
\newblock Partial success in closing the gap between human and machine vision.
\newblock {\em NeurIPS}.

\bibitem[Gifford et~al., 2022]{Gifford2022}
Gifford, A.~T., Dwivedi, K., Roig, G., and Cichy, R.~M. (2022).
\newblock {A large and rich EEG dataset for modeling human visual object recognition}.
\newblock {\em NeuroImage}, 264:119754.

\bibitem[G{\"{u}}{\c{c}}l{\"{u}} and van Gerven, 2015]{Guclu2015}
G{\"{u}}{\c{c}}l{\"{u}}, U. and van Gerven, M.~A. (2015).
\newblock {Deep Neural Networks Reveal a Gradient in the Complexity of Neural Representations across the Ventral Stream}.
\newblock {\em Journal of Neuroscience}, 35(27):10005--10014.

\bibitem[Han and Sereno, 2022]{Han2022}
Han, Z. and Sereno, A. (2022).
\newblock {Modeling the Ventral and Dorsal Cortical Visual Pathways Using Artificial Neural Networks}.
\newblock {\em Neural Computation}, 34(1):138--171.

\bibitem[Han and Sereno, 2023]{Han2023}
Han, Z. and Sereno, A. (2023).
\newblock {Identifying and Localizing Multiple Objects Using Artificial Ventral and Dorsal Cortical Visual Pathways}.
\newblock {\em Neural Computation}, 35(2):249--275.

\bibitem[Hebart et~al., 2019]{Hebart2019}
Hebart, M.~N., Dickter, A.~H., Kidder, A., Kwok, W.~Y., Corriveau, A., {Van Wicklin}, C., and Baker, C.~I. (2019).
\newblock {THINGS: A database of 1,854 object concepts and more than 26,000 naturalistic object images}.
\newblock {\em PLoS ONE}, 14(10):1--24.

\bibitem[Hebart et~al., 2020]{Hebart2020}
Hebart, M.~N., Zheng, C.~Y., Pereira, F., and Baker, C.~I. (2020).
\newblock {Revealing the multidimensional mental representations of natural objects underlying human similarity judgements}.
\newblock {\em Nature Human Behaviour}, 4(11):1173--1185.

\bibitem[Horikawa and Kamitani, 2017]{Horikawa2017}
Horikawa, T. and Kamitani, Y. (2017).
\newblock {Generic decoding of seen and imagined objects using hierarchical visual features}.
\newblock {\em Nature Communications}, 8:15037.

\bibitem[Jain et~al., 2023]{Jain2023}
Jain, N., Wang, A., Henderson, M.~M., Lin, R., Prince, J.~S., Tarr, M.~J., and Wehbe, L. (2023).
\newblock {Selectivity for food in human ventral visual cortex}.
\newblock {\em Communications Biology}, 6(1):1--14.

\bibitem[Khaligh-Razavi et~al., 2018]{Khaligh-Razavi2018}
Khaligh-Razavi, S.-M., Cichy, R.~M., Pantazis, D., and Oliva, A. (2018).
\newblock {Tracking the Spatiotemporal Neural Dynamics of Real-world Object Size and Animacy in the Human Brain}.
\newblock {\em Journal of Cognitive Neuroscience}, 30(11):1559--1576.

\bibitem[Khosla et~al., 2022]{Khosla2022}
Khosla, M., {Ratan Murty}, N.~A., and Kanwisher, N. (2022).
\newblock {A highly selective response to food in human visual cortex revealed by hypothesis-free voxel decomposition}.
\newblock {\em Current Biology}, 32(19):4159--4171.e9.

\bibitem[Kietzmann et~al., 2019]{Kietzmann2019}
Kietzmann, T.~C., Spoerer, C.~J., S{\"{o}}rensen, L.~K., Cichy, R.~M., Hauk, O., and Kriegeskorte, N. (2019).
\newblock {Recurrence is required to capture the representational dynamics of the human visual system}.
\newblock {\em Proceedings of the National Academy of Sciences of the United States of America}, 116(43):21854--21863.

\bibitem[Konkle and Alvarez, 2023]{Konkle2023}
Konkle, T. and Alvarez, G. (2023).
\newblock {Cognitive Steering in Deep Neural Networks via Long-Range Modulatory Feedback Connections}.
\newblock {\em Advances in Neural Information Processing Systems (NeurIPS)}.

\bibitem[Konkle and Alvarez, 2022]{Konkle2022}
Konkle, T. and Alvarez, G.~A. (2022).
\newblock {A self-supervised domain-general learning framework for human ventral stream representation}.
\newblock {\em Nature Communications}, 13(1):1--12.

\bibitem[Konkle and Caramazza, 2013]{Konkle2013}
Konkle, T. and Caramazza, A. (2013).
\newblock {Tripartite Organization of the Ventral Stream by Animacy and Object Size}.
\newblock {\em Journal of Neuroscience}, 33(25):10235--10242.

\bibitem[Kriegeskorte et~al., 2008]{Kriegeskorte2008a}
Kriegeskorte, N., Mur, M., and Bandettini, P. (2008).
\newblock {Representational similarity analysis - connecting the branches of systems neuroscience}.
\newblock {\em Frontiers in Systems Neuroscience}, 4.

\bibitem[Kubilius et~al., 2019]{Kubilius2019}
Kubilius, J., Schrimpf, M., Kar, K., Rajalingham, R., Hong, H., Majaj, N.~J., Issa, E.~B., Bashivan, P., Prescott-Roy, J., Schmidt, K., Nayebi, A., Bear, D., Yamins, D. L.~K., and Dicarlo, J.~J. (2019).
\newblock {Brain-Like Object Recognition with High-Performing Shallow Recurrent ANNs}.
\newblock {\em Advances in Neural Information Processing Systems (NeurIPS)}, 32.

\bibitem[Kubilius et~al., 2018]{Kubilius2018}
Kubilius, J., Schrimpf, M., Nayebi, A., Bear, D., Yamins, D., and DiCarlo, J. (2018).
\newblock {CORnet: Modeling the Neural Mechanisms of Core Object Recognition}.
\newblock {\em bioRxiv}.

\bibitem[Lecun et~al., 2015]{Lecun2015}
Lecun, Y., Bengio, Y., and Hinton, G. (2015).
\newblock {Deep learning}.
\newblock {\em Nature}, 521(7553).

\bibitem[Lee et~al., 2020]{Lee2020}
Lee, H., Margalit, E., Jozwik, K.~M., Cohen, M.~A., Kanwisher, N., Yamins, D. L.~K., and DiCarlo, J.~J. (2020).
\newblock {Topographic deep artificial neural networks reproduce the hallmarks of the primate inferior temporal cortex face processing network}.
\newblock {\em bioRxiv}.

\bibitem[Li et~al., 2019]{Li2019}
Li, Z., Brendel, W., Walker, E.~Y., Cobos, E., Muhammad, T., Reimer, J., Bethge, M., Sinz, F.~H., Pitkow, X., and Tolias, A.~S. (2019).
\newblock {Learning from brains how to regularize machines}.
\newblock {\em Advances in Neural Information Processing Systems (NeurIPS)}, 32.

\bibitem[Lu et~al., 2023]{Lu2023e}
Lu, Z., Doerig, A., Bosch, V., Krahmer, B., Kaiser, D., Cichy, R.~M., and Kietzmann, T.~C. (2023).
\newblock {End-to-end topographic networks as models of cortical map formation and human visual behaviour: moving beyond convolutions}.
\newblock {\em arXiv}.

\bibitem[Lu and Golomb, 2023a]{Lu2023c}
Lu, Z. and Golomb, J.~D. (2023a).
\newblock {Generate your neural signals from mine: individual-to-individual EEG converters}.
\newblock {\em Proceedings of the Annual Meeting of the Cognitive Science Society 45}.

\bibitem[Lu and Golomb, 2023b]{Lu2023d}
Lu, Z. and Golomb, J.~D. (2023b).
\newblock {Human EEG and artificial neural networks reveal disentangled representations of object real-world size in natural images}.
\newblock {\em bioRxiv}.

\bibitem[Lu and Ku, 2020]{Lu2020}
Lu, Z. and Ku, Y. (2020).
\newblock {NeuroRA: A Python Toolbox of Representational Analysis From Multi-Modal Neural Data}.
\newblock {\em Frontiers in Neuroinformatics}, 14:61.

\bibitem[Lu et~al., 2024]{Lu2024}
Lu, Z., Wang, Y., and Golomb, J.~D. (2024).
\newblock {ReAlnet: Achieving More Human Brain-Like Vision via Human Neural Representational Alignment}.
\newblock {\em arXiv}.

\bibitem[Margalit et~al., 2023]{Margalit2023}
Margalit, E., Lee, H., Finzi, D., DiCarlo, J.~J., Grill-Spector, K., and Yamins, D.~L. (2023).
\newblock {A Unifying Principle for the Functional Organization of Visual Cortex}.
\newblock {\em bioRxiv}.

\bibitem[Pirlot et~al., 2022]{Pirlot2022}
Pirlot, C., Gerum, R.~C., Efird, C., Zylberberg, J., and Fyshe, A. (2022).
\newblock {Improving the Accuracy and Robustness of CNNs Using a Deep CCA Neural Data Regularizer}.
\newblock {\em arXiv}.

\bibitem[Prince et~al., 2023]{Prince2023}
Prince, J.~S., Alvarez, G.~A., and Konkle, T. (2023).
\newblock {A contrastive coding account of category selectivity in the ventral visual stream}.
\newblock {\em bioRxiv}.

\bibitem[Rajalingham et~al., 2018]{Rajalingham2018}
Rajalingham, R., Issa, E.~B., Bashivan, P., Kar, K., Schmidt, K., and DiCarlo, J.~J. (2018).
\newblock {Large-Scale, High-Resolution Comparison of the Core Visual Object Recognition Behavior of Humans, Monkeys, and State-of-the-Art Deep Artificial Neural Networks}.
\newblock {\em Journal of Neuroscience}, 38(33):7255--7269.

\bibitem[Safarani et~al., 2021]{Safarani2021}
Safarani, S., Nix, A., Willeke, K., Cadena, S.~A., Restivo, K., Denfield, G., Tolias, A.~S., and Sinz, F.~H. (2021).
\newblock {Towards robust vision by multi-task learning on monkey visual cortex}.
\newblock {\em Advances in Neural Information Processing Systems (NeurIPS)}, 34.

\bibitem[Schrimpf et~al., 2020]{Schrimpf2020}
Schrimpf, M., Kubilius, J., Hong, H., Majaj, N.~J., Rajalingham, R., Issa, E.~B., Kar, K., Bashivan, P., Prescott-Roy, J., Geiger, F., Schmidt, K., Yamins, D. L.~K., and DiCarlo, J.~J. (2020).
\newblock Brain-{Score}: {Which} {Artificial} {Neural} {Network} for {Object} {Recognition} is most {Brain}-{Like}?
\newblock {\em bioRxiv}.

\bibitem[Shao et~al., 2024]{Shao2024}
Shao, Z., Ma, L., Li, B., and Beck, D.~M. (2024).
\newblock {Leveraging the Human Ventral Visual Stream to Improve Neural Network Robustness}.
\newblock {\em arXiv}.

\bibitem[Shen et~al., 2019]{Shen2019}
Shen, G., Horikawa, T., Majima, K., and Kamitani, Y. (2019).
\newblock {Deep image reconstruction from human brain activity}.
\newblock {\em PLoS computational biology}, 15(1):e1006633.

\bibitem[Sun et~al., 2017]{Sun2017}
Sun, T., Wang, Y., Yang, J., and Hu, X. (2017).
\newblock {Convolution Neural Networks With Two Pathways for Image Style Recognition}.
\newblock {\em IEEE Transactions on Image Processing}, 26(9):4102--4113.

\bibitem[Yamins et~al., 2014]{Yamins2014}
Yamins, D.~L., Hong, H., Cadieu, C.~F., Solomon, E.~A., Seibert, D., and DiCarlo, J.~J. (2014).
\newblock {Performance-optimized hierarchical models predict neural responses in higher visual cortex}.
\newblock {\em Proceedings of the National Academy of Sciences of the United States of America}, 111(23):8619--8624.

\bibitem[Yargholi and de~Beeck, 2023]{Yargholi2023}
Yargholi, E. and de~Beeck, H.~O. (2023).
\newblock {Category Trumps Shape as an Organizational Principle of Object Space in the Human Occipitotemporal Cortex}.
\newblock {\em Journal of Neuroscience}, 43(16):2960--2972.

\bibitem[Yosinski et~al., 2015]{Yosinski2015}
Yosinski, J., Clune, J., Nguyen, A., Fuchs, T., and Lipson, H. (2015).
\newblock {Understanding Neural Networks Through Deep Visualization}.
\newblock {\em arXiv}.

\bibitem[Zeiler and Fergus, 2014]{Zeiler2014}
Zeiler, M. and Fergus, R. (2014).
\newblock {Visualizing and Understanding Convolutional Networks}.
\newblock {\em Proceedings of 13th European Conference on Computer Vision}, 8689:808--833.

\end{thebibliography}

\newpage

\section*{Supplementary}

\setcounter{figure}{0}
\renewcommand{\thefigure}{S\arabic{figure}}

\begin{figure}[h!]
\begin{center}
\centerline{\includegraphics[width=0.8\columnwidth]{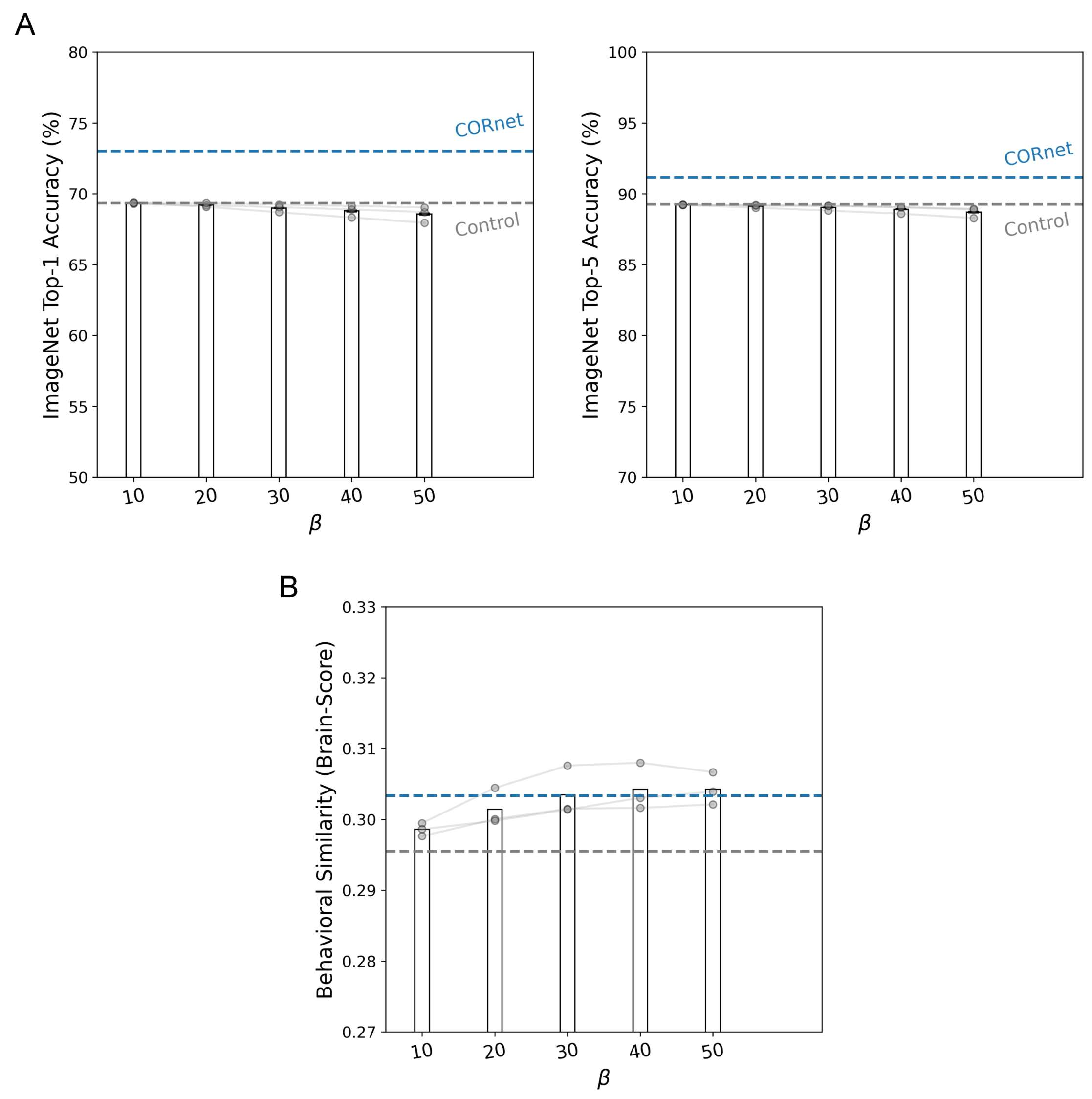}}
\caption{(A) Classification accuracy on ImageNet and (B) behavior similarity on Brain-Score of ReAlnet-fMRIs at different $\beta$ values.}
\label{FigureS1}
\end{center}
\end{figure}

\begin{figure}[h!]
\begin{center}
\centerline{\includegraphics[width=0.8\columnwidth]{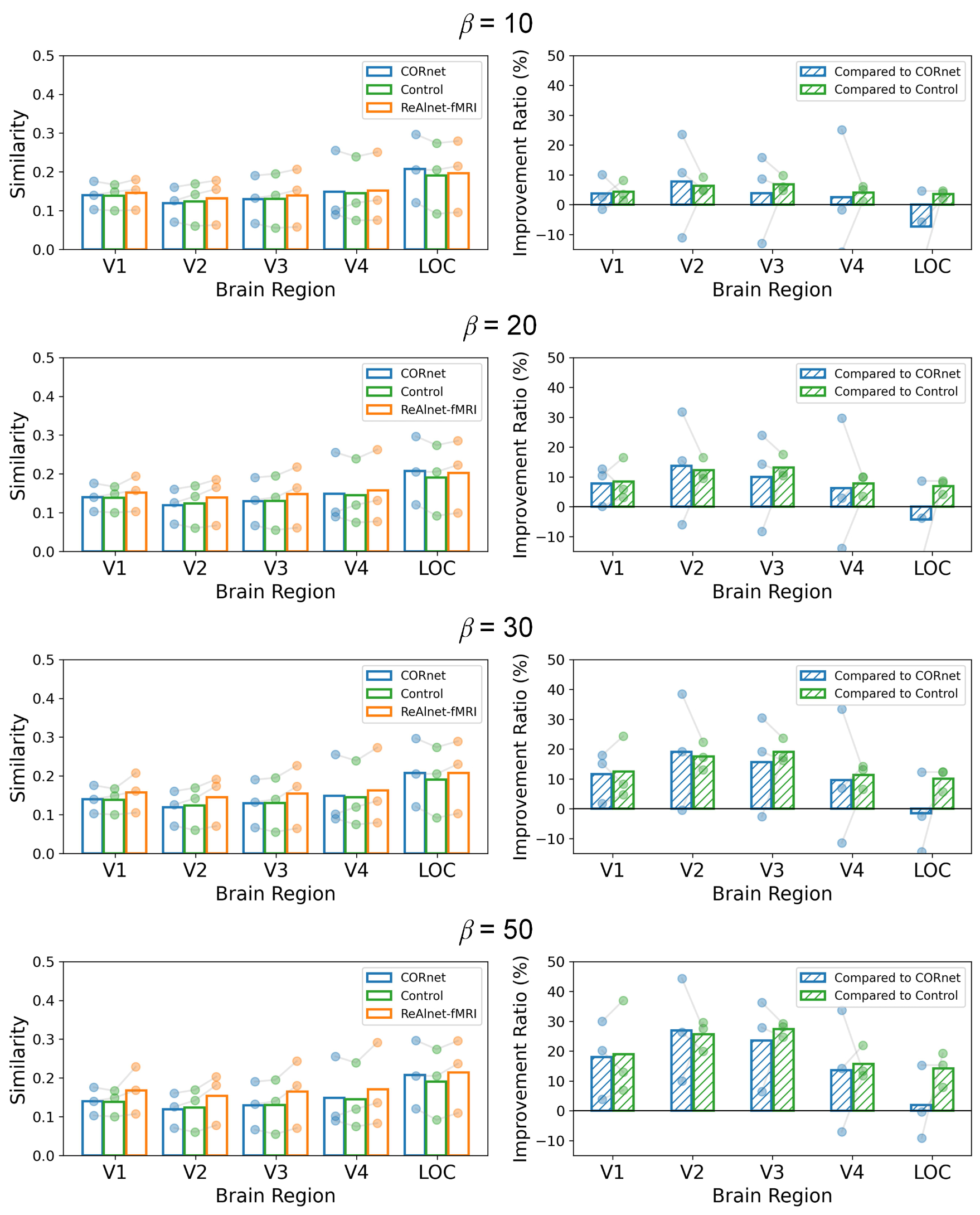}}
\caption{Within-subject model-fMRI similarity and similarity improvement ratio on natural images of ReAlnet-fMRIs with $\beta$ = 10, 20, 30, and 50. Each circle dot indicates an individual ReAlnet-fMRI.}
\label{FigureS2}
\end{center}
\end{figure}

\begin{figure}[h!]
\begin{center}
\centerline{\includegraphics[width=0.8\columnwidth]{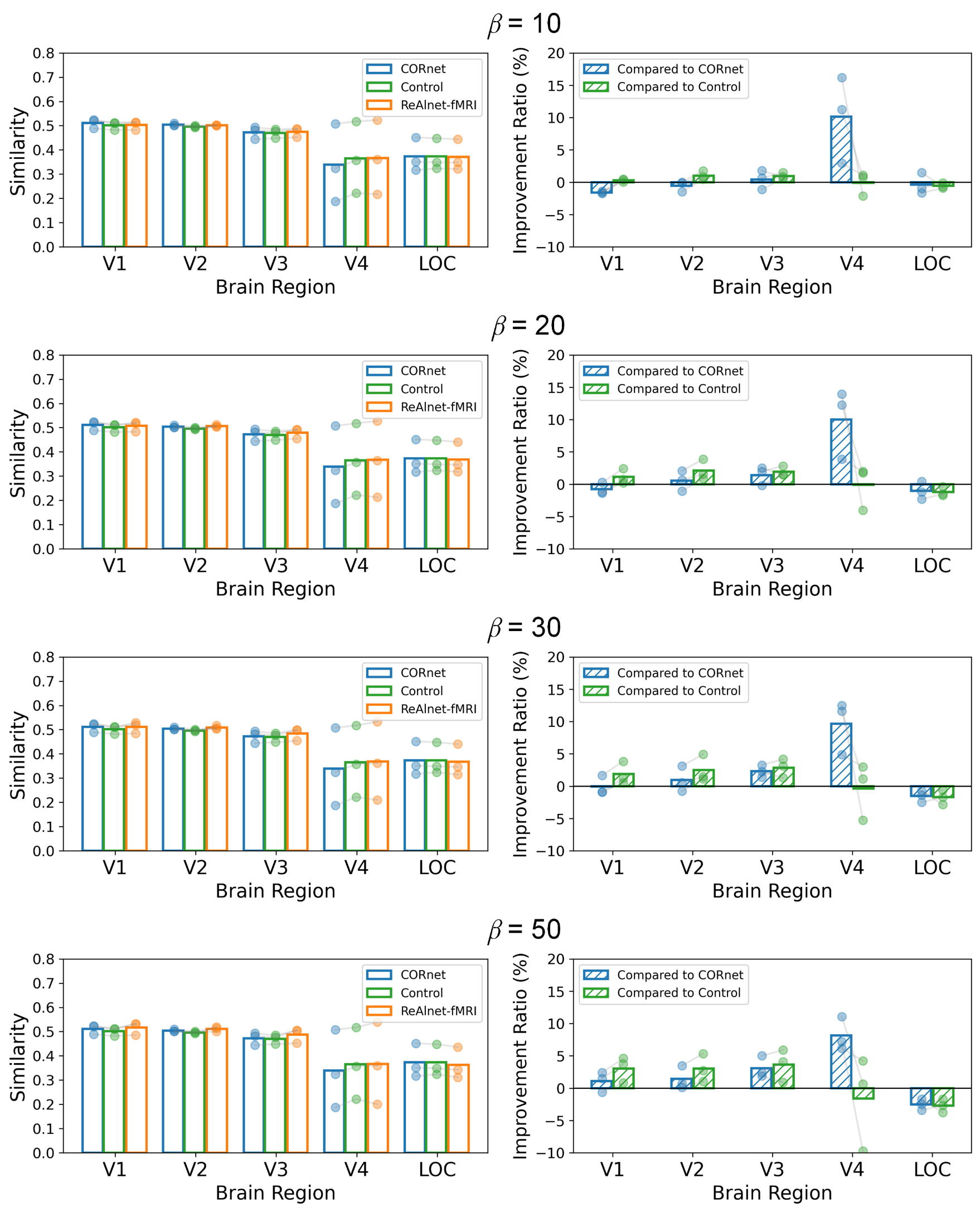}}
\caption{Within-subject model-fMRI similarity and similarity improvement ratio on artificial shape images of ReAlnet-fMRIs with $\beta$ = 10, 20, 30, and 50. Each circle dot indicates an individual ReAlnet-fMRI.}
\label{FigureS3}
\end{center}
\end{figure}

\begin{figure}[h!]
\begin{center}
\centerline{\includegraphics[width=0.8\columnwidth]{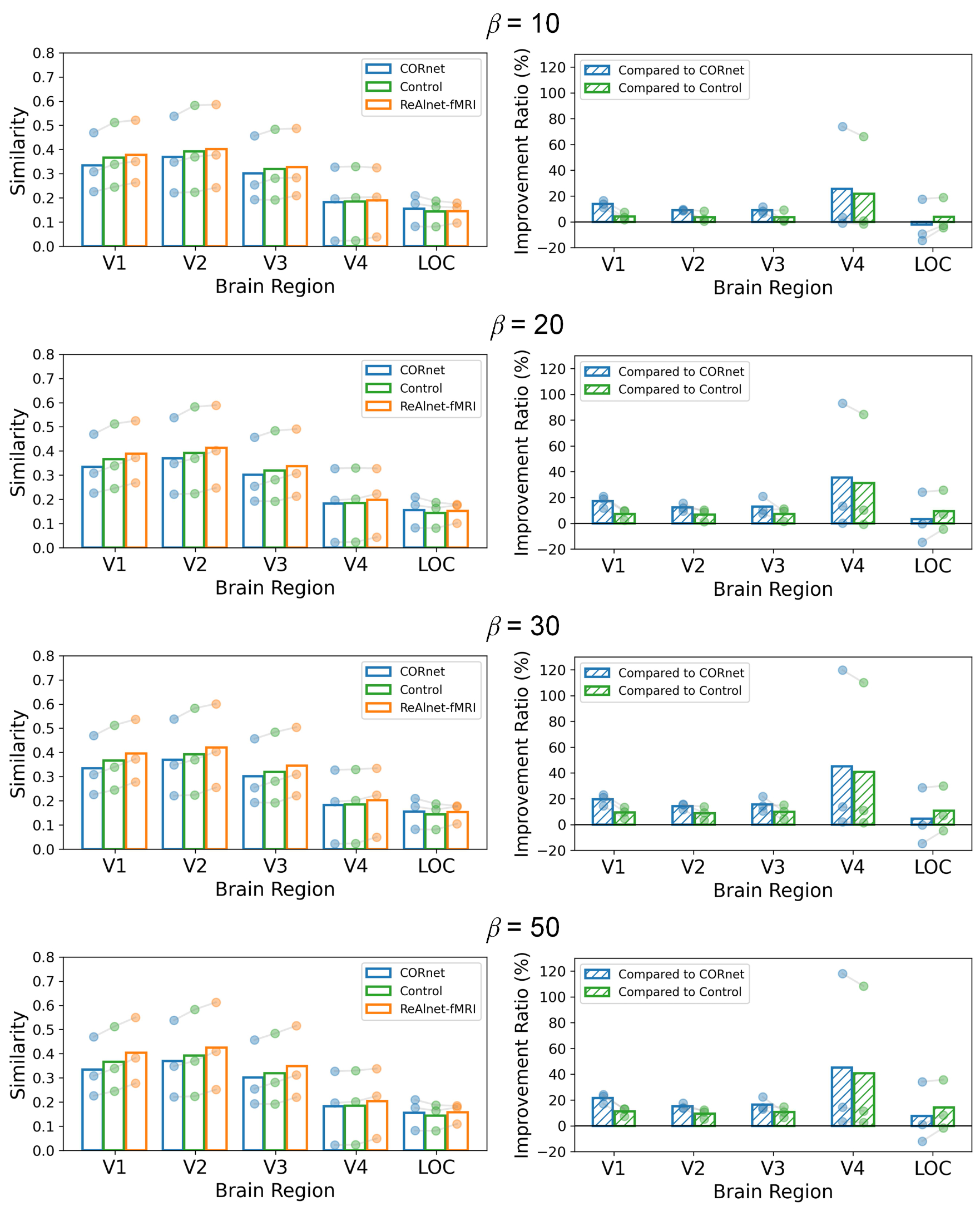}}
\caption{Within-subject model-fMRI similarity and similarity improvement ratio on alphabetical letter images of ReAlnet-fMRIs with $\beta$ = 10, 20, 30, and 50. Each circle dot indicates an individual ReAlnet-fMRI.}
\label{FigureS4}
\end{center}
\end{figure}

\begin{figure}[h!]
\begin{center}
\centerline{\includegraphics[width=\columnwidth]{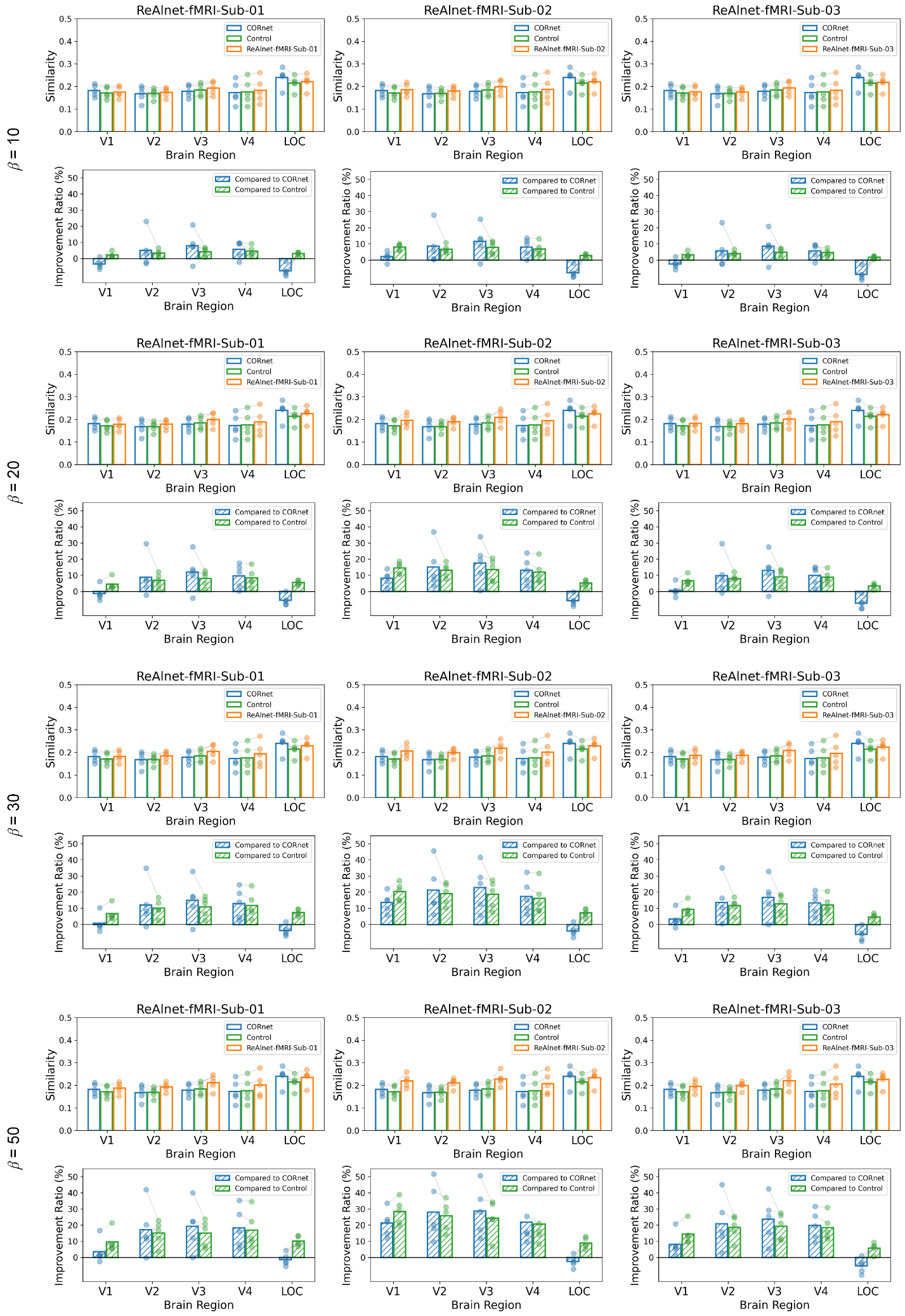}}
\caption{Across-subject model-fMRI similarity and similarity improvement ratio of ReAlnet-fMRIs with $\beta$ = 10, 20, 30, and 50. Each circle dot indicates a subject from \textit{Horikawa fMRI dataset}.}
\label{FigureS5}
\end{center}
\end{figure}

\begin{figure}[h!]
\begin{center}
\centerline{\includegraphics[width=\columnwidth]{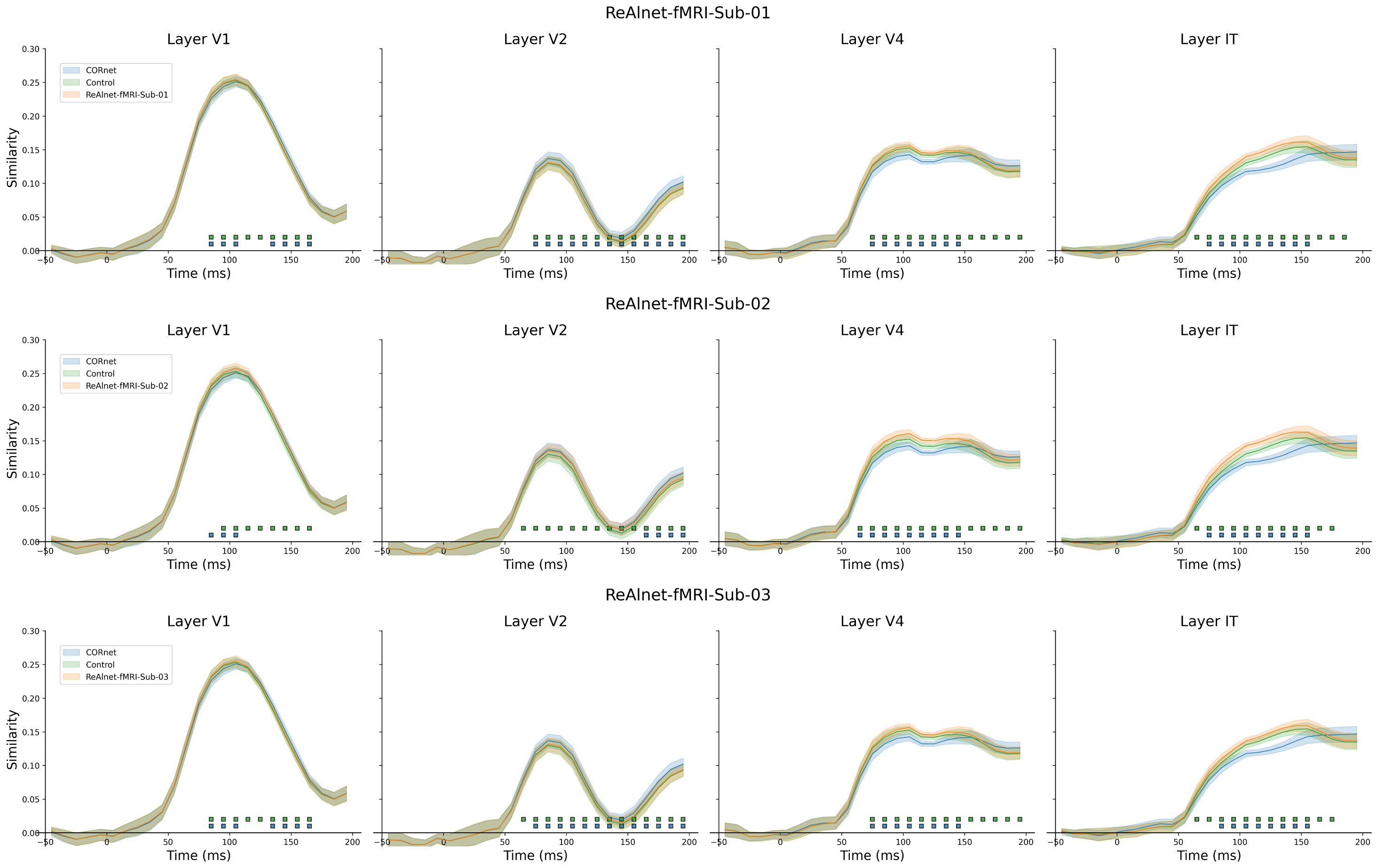}}
\caption{Across-subject temporal model-EEG similarity of ReAlnet-fMRIs with $\beta$ = 10. Blue and green square dots with black outlines at the bottom indicate the timepoints where ReAlnet-fMRI vs. CORnet and ReAlnet-fMRI vs. Control were significantly different ($p<.05$). Shaded area reflects ±SEM.}
\label{FigureS6}
\end{center}
\end{figure}

\begin{figure}[h!]
\begin{center}
\centerline{\includegraphics[width=\columnwidth]{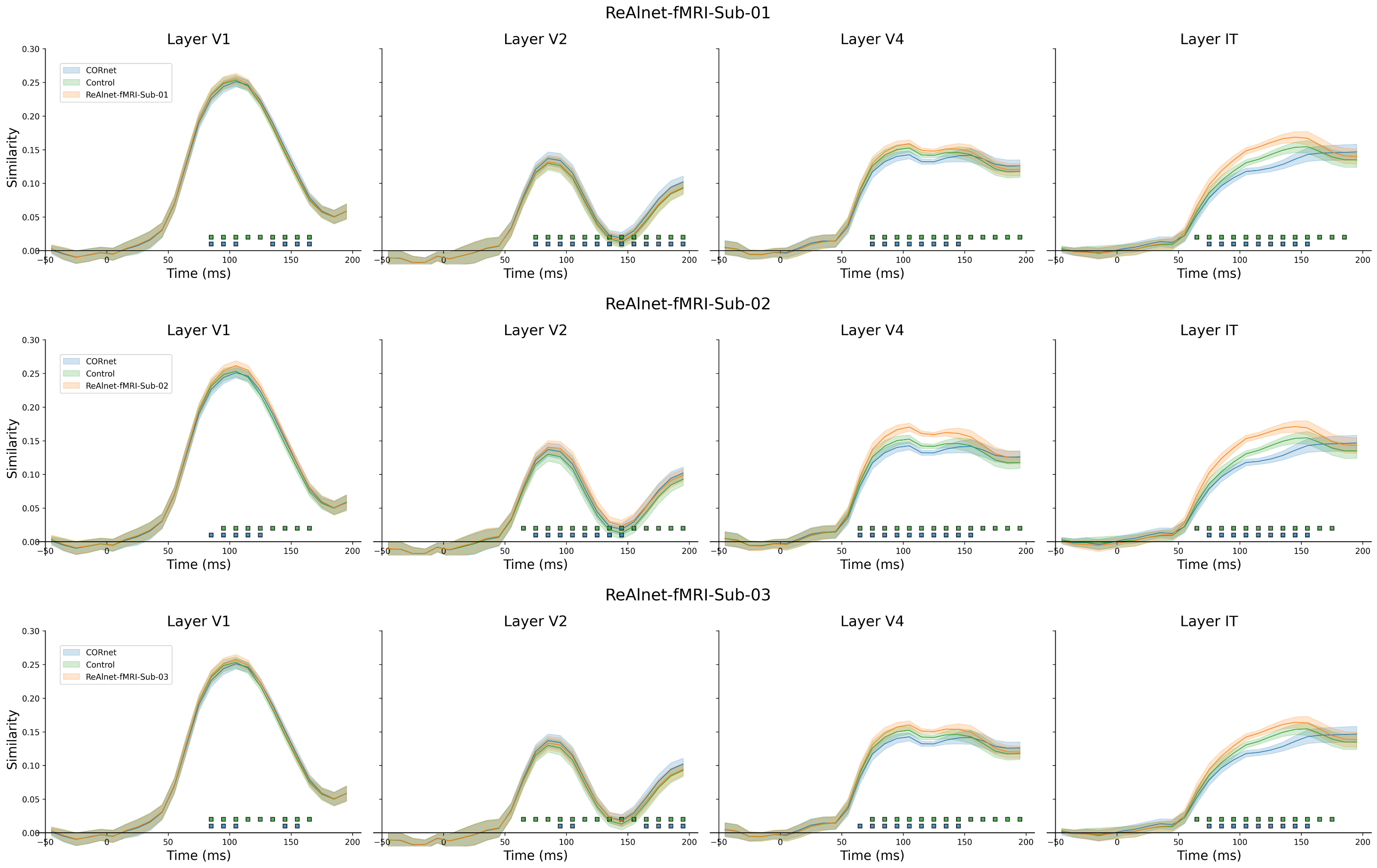}}
\caption{Across-subject temporal model-EEG similarity of ReAlnet-fMRIs with $\beta$ = 20. Blue and green square dots with black outlines at the bottom indicate the timepoints where ReAlnet-fMRI vs. CORnet and ReAlnet-fMRI vs. Control were significantly different ($p<.05$). Shaded area reflects ±SEM.}
\label{FigureS7}
\end{center}
\end{figure}

\begin{figure}[h!]
\begin{center}
\centerline{\includegraphics[width=\columnwidth]{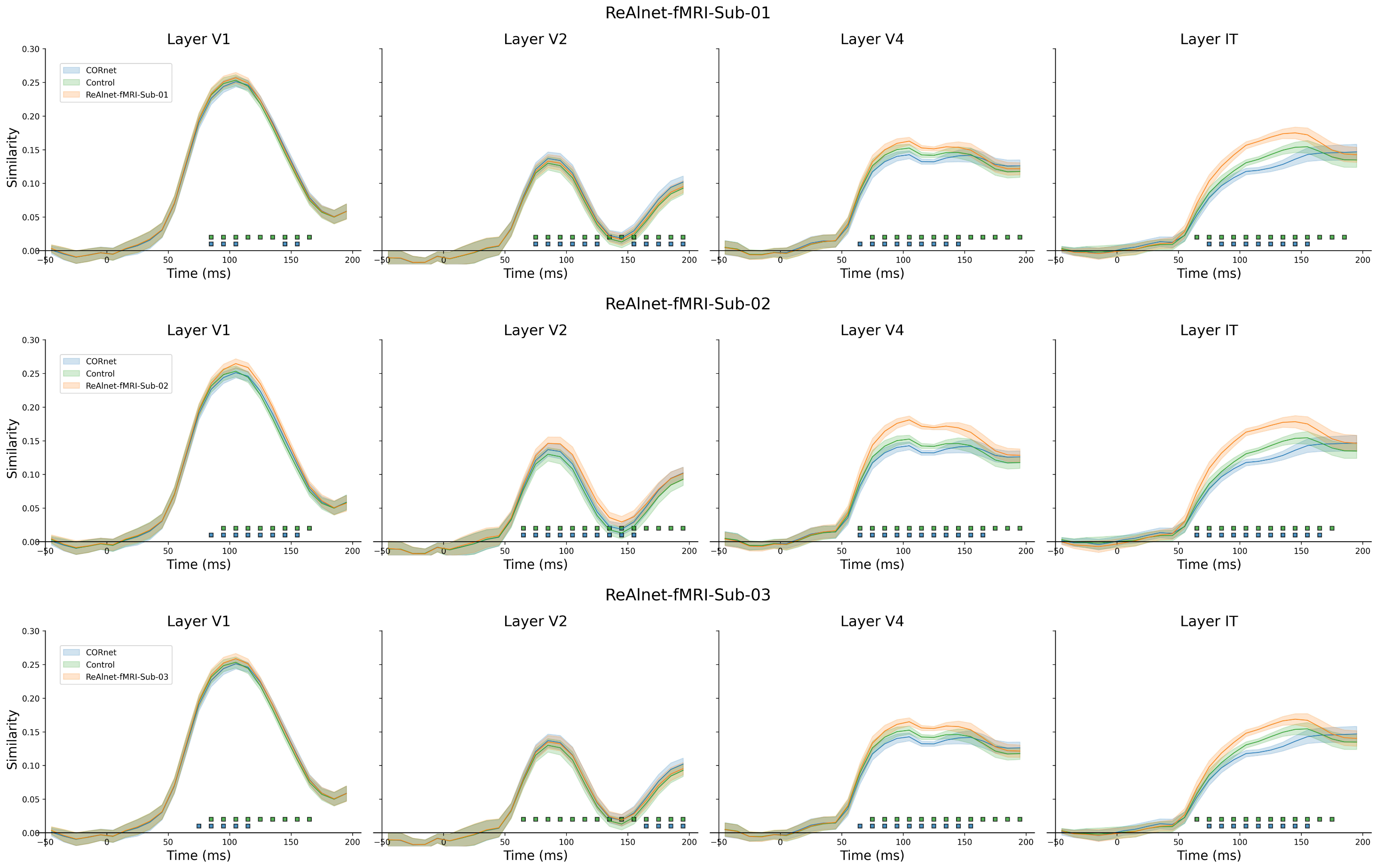}}
\caption{Across-subject temporal model-EEG similarity of ReAlnet-fMRIs with $\beta$ = 30. Blue and green square dots with black outlines at the bottom indicate the timepoints where ReAlnet-fMRI vs. CORnet and ReAlnet-fMRI vs. Control were significantly different ($p<.05$). Shaded area reflects ±SEM.}
\label{FigureS8}
\end{center}
\end{figure}

\begin{figure}[h!]
\begin{center}
\centerline{\includegraphics[width=\columnwidth]{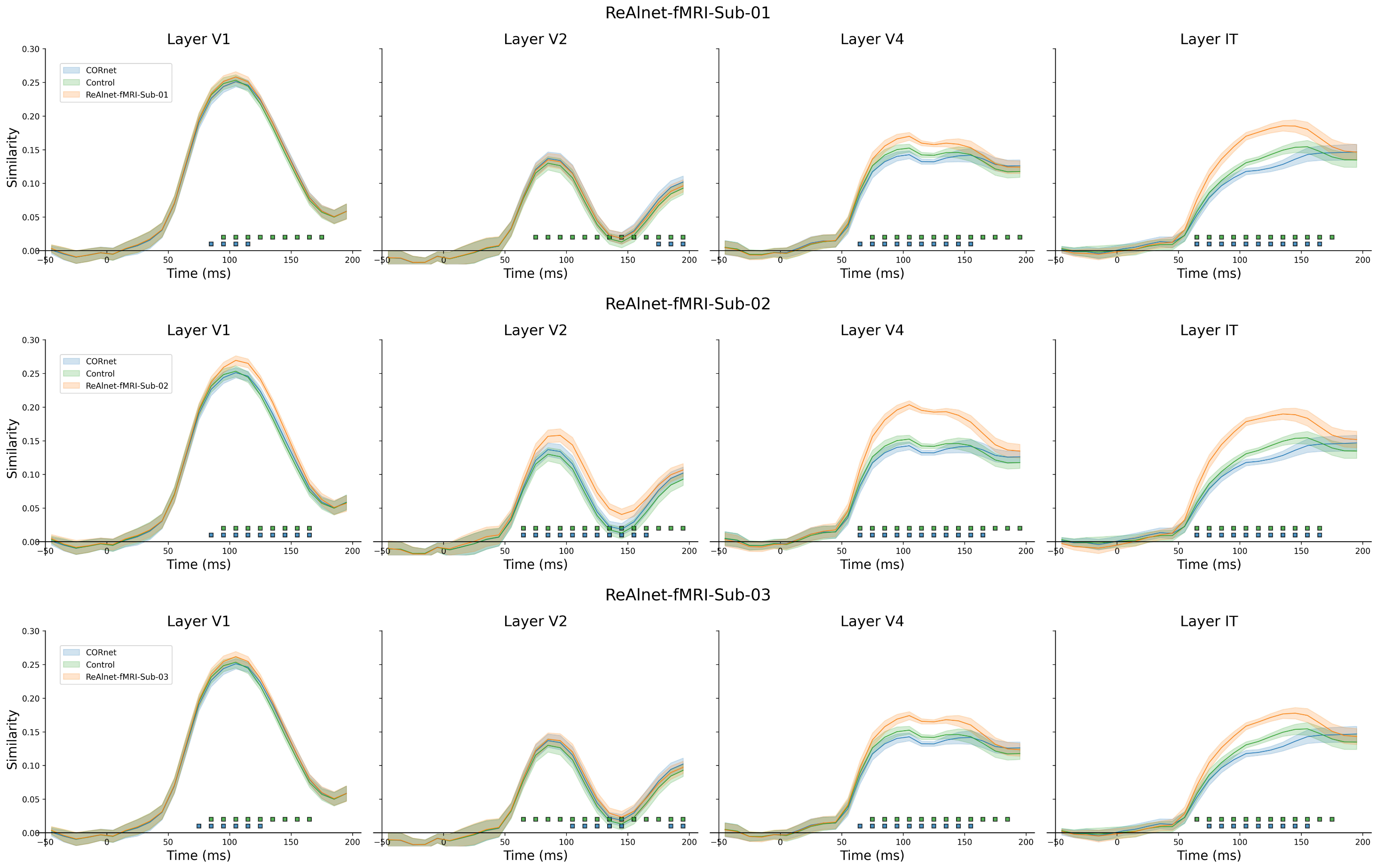}}
\caption{Across-subject temporal model-EEG similarity of ReAlnet-fMRIs with $\beta$ = 50. Blue and green square dots with black outlines at the bottom indicate the timepoints where ReAlnet-fMRI vs. CORnet and ReAlnet-fMRI vs. Control were significantly different ($p<.05$). Shaded area reflects ±SEM.}
\label{FigureS9}
\end{center}
\end{figure}

\begin{figure}[h!]
\begin{center}
\centerline{\includegraphics[width=0.9\columnwidth]{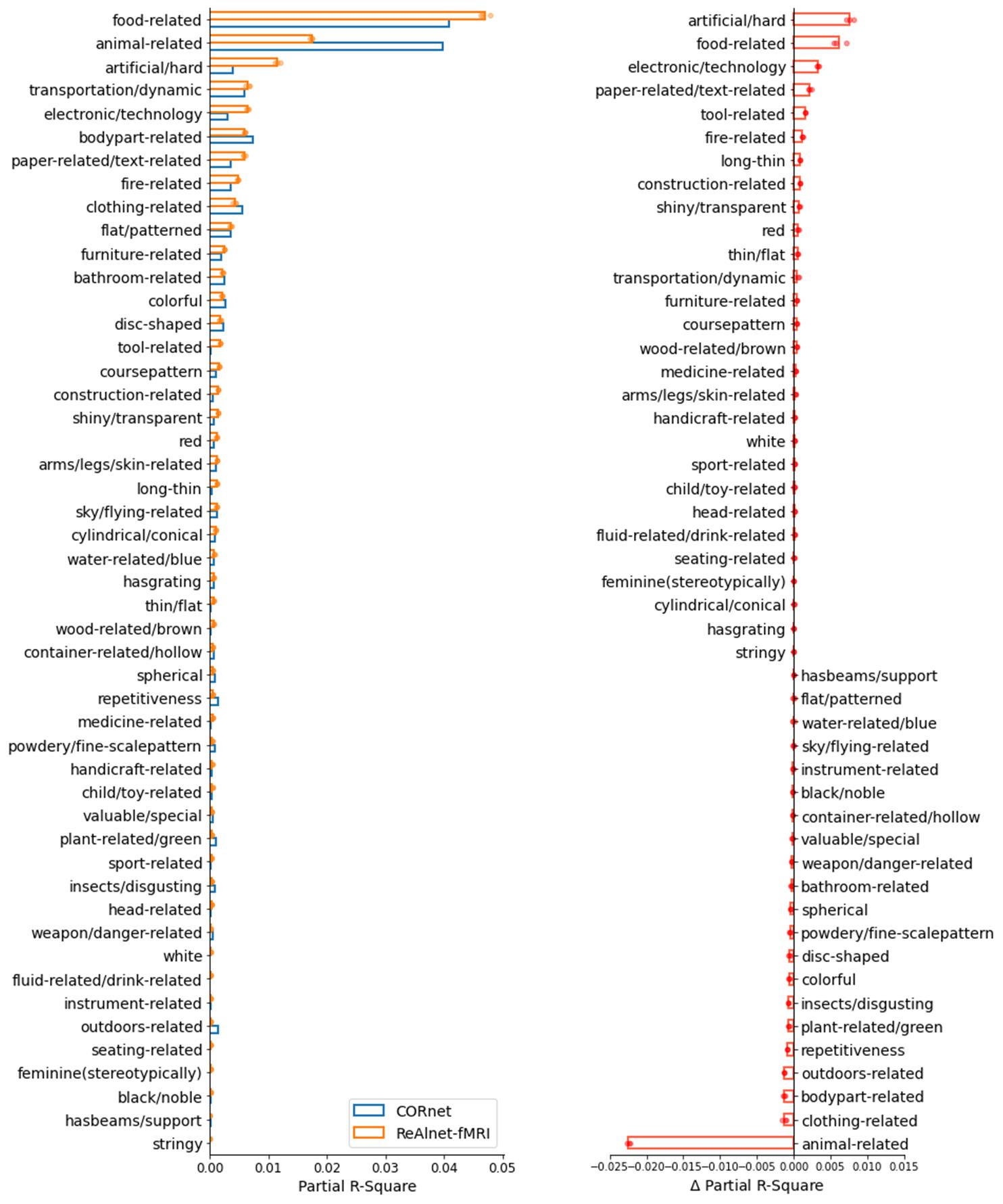}}
\caption{Internal representations in ReAlnet-fMRIs with $\beta$ = 10 and CORnet. Each circle dot indicates an individual ReAlnet-fMRI.}
\label{FigureS10}
\end{center}
\end{figure}

\begin{figure}[h!]
\begin{center}
\centerline{\includegraphics[width=0.9\columnwidth]{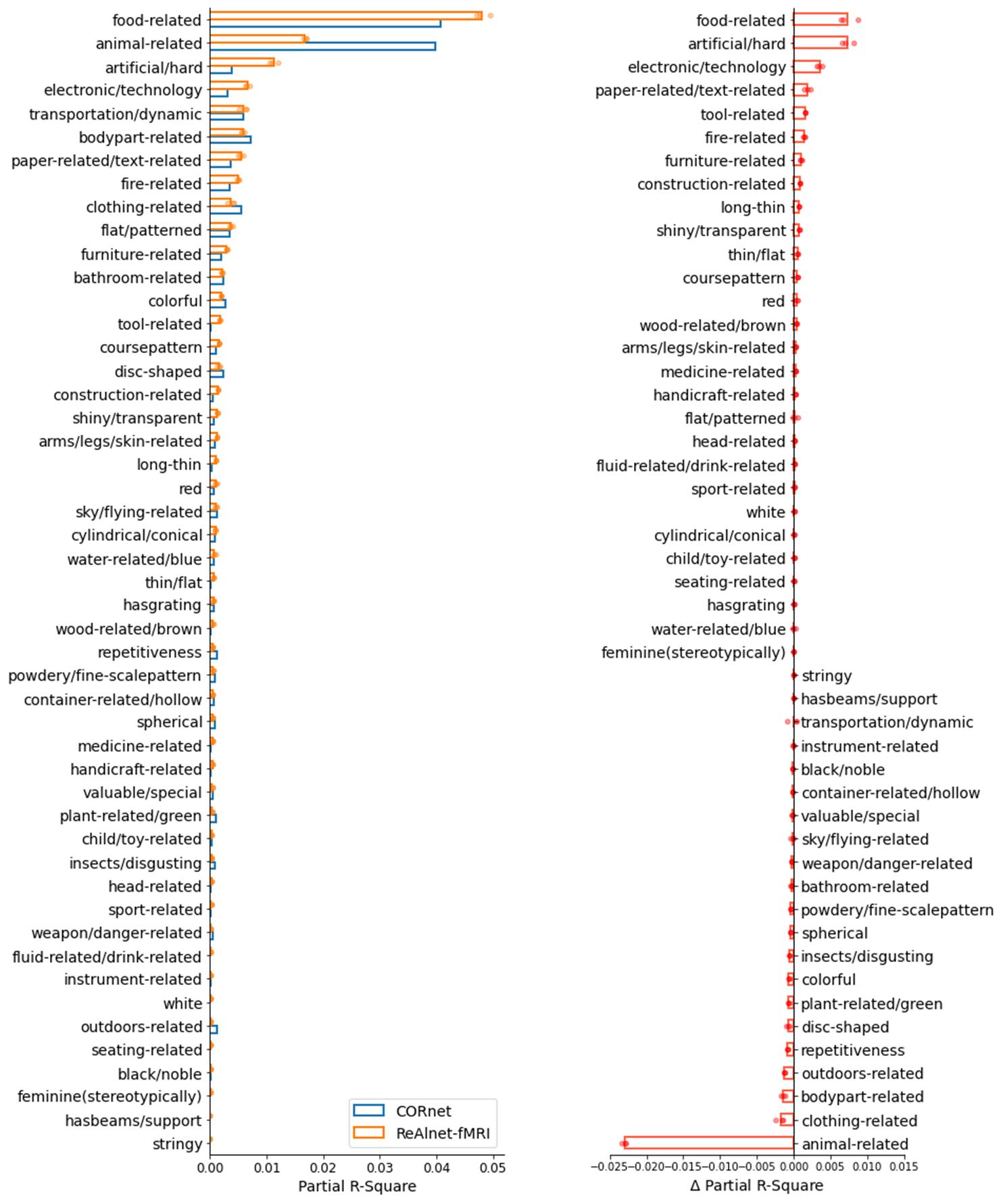}}
\caption{Internal representations in ReAlnet-fMRIs with $\beta$ = 20 and CORnet. Each circle dot indicates an individual ReAlnet-fMRI.}
\label{FigureS11}
\end{center}
\end{figure}

\begin{figure}[h!]
\begin{center}
\centerline{\includegraphics[width=0.9\columnwidth]{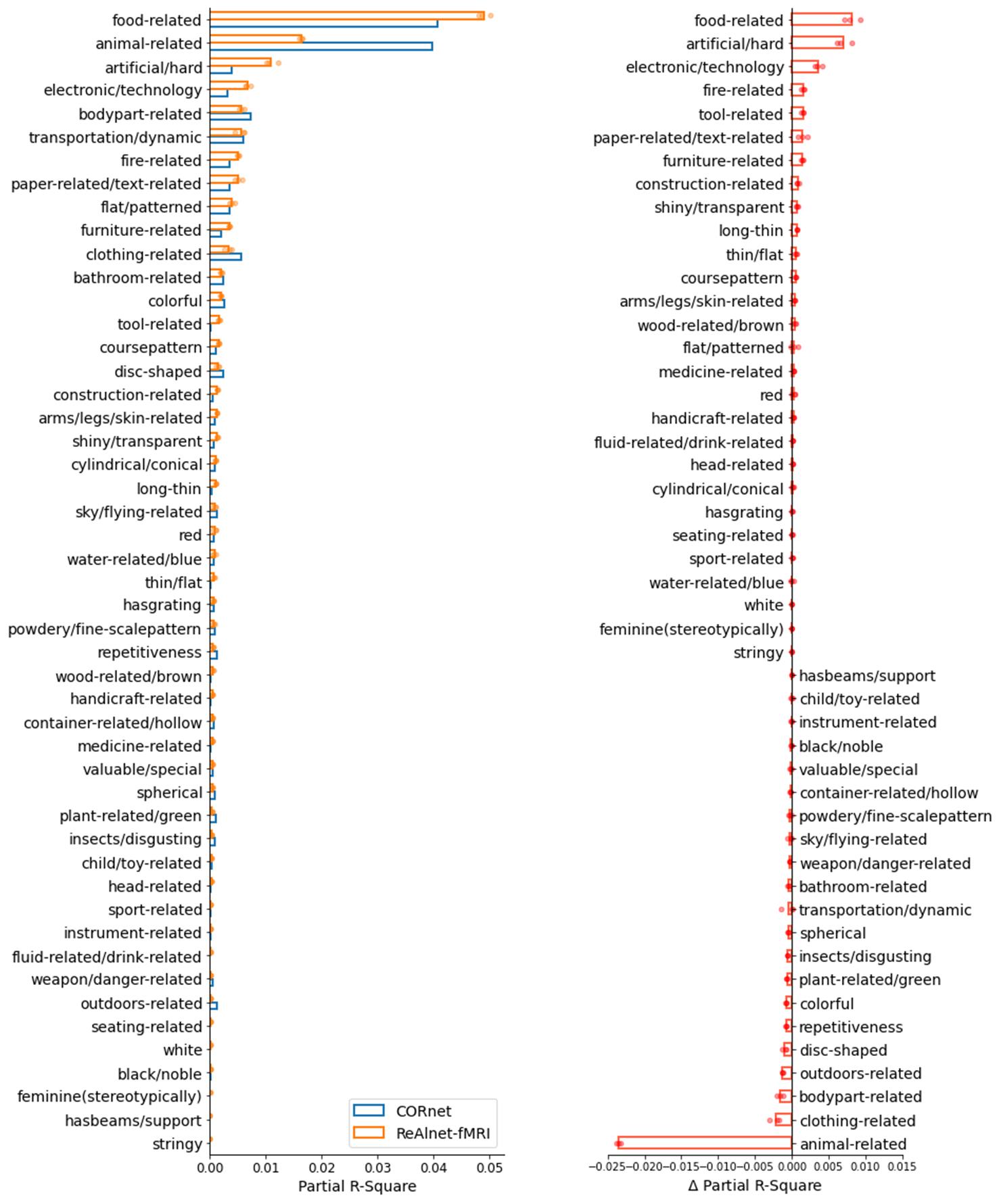}}
\caption{Internal representations in ReAlnet-fMRIs with $\beta$ = 30 and CORnet. Each circle dot indicates an individual ReAlnet-fMRI.}
\label{FigureS12}
\end{center}
\end{figure}

\begin{figure}[h!]
\begin{center}
\centerline{\includegraphics[width=0.9\columnwidth]{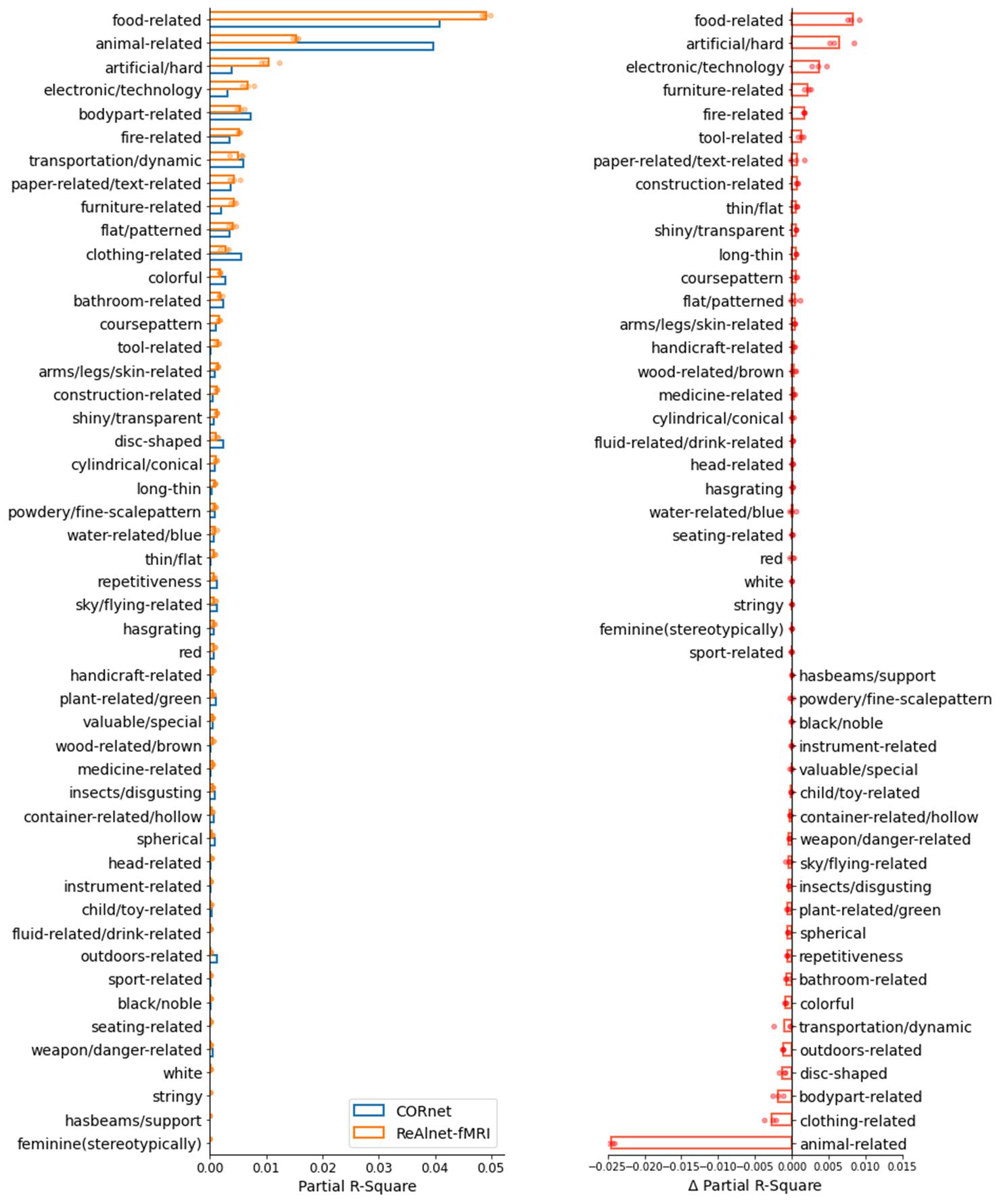}}
\caption{Internal representations in ReAlnet-fMRIs with $\beta$ = 50 and CORnet. Each circle dot indicates an individual ReAlnet-fMRI.}
\label{FigureS13}
\end{center}
\end{figure}

\end{document}